\documentclass{aa}

\usepackage{natbib}
\usepackage{graphicx}
\usepackage{epsfig}
\usepackage{subfigure}
\usepackage{amssymb}
\usepackage{longtable}
\usepackage{rotate}
\usepackage{rotating}
\usepackage{supertabular}
\usepackage{lscape}
\def\wings{{\sc wings}}
\def\sdss{{\sc sdss}}
\begin{document}

\title{WINGS-SPE II:\\
A catalog of stellar ages and star formation histories, stellar masses and dust extinction values for local clusters galaxies}

\author{J. Fritz\inst{1,2} \and B.~M. Poggianti\inst{1} \and A. Cava\inst{3,4} \and T. Valentinuzzi\inst{5} 
\and A. Moretti\inst{1} \and D. Bettoni\inst{1} \and A. Bressan\inst{1} \and W.~J. Couch\inst{6} \and M. D'Onofrio\inst{5} 
\and A.~Dressler\inst{7} \and G. Fasano\inst{1} \and P. Kj\ae rgaard\inst{8} \and M. Moles\inst{9}  \and A. Omizzolo\inst{1,10} 
\and J. Varela\inst{1}}
\offprints{Jacopo Fritz,\\ 
           \email{jacopo.fritz@UGent.be}}
\institute{INAF-Osservatorio Astronomico di Padova, vicolo Osservatorio 5, 35122 Padova, Italy\\
\email{jacopo.fritz@UGent.be}
\and
Sterrenkundig Observatorium Vakgroep Fysica en Sterrenkunde Universeit Gent, Krijgslaan 281, S9  9000 Gent
\and 
 Instituto de Astrof\'\i sica de Canarias, E-38200 La Laguna, Tenerife, Spain
\and
Departamento de Astrof\'\i sica, Universidad de La Laguna, E-38205 La Laguna, Tenerife, Spain
\and
 Dipartimento di Astronomia, vicolo Osservatorio 2, 35122 Padova, Italy
\and
Centre for Astrophysics and Supercomputing, Swinburne University of Technology, Melbourne, Australia
\and
 Observatories of the Carnegie Institution of Washington, Pasadena, CA 91101, USA
\and
 Copenhagen University Observatory. The Niels Bohr Insitute for Astronomy Physics and Geophysics, Juliane Maries Vej 30, 2100 Copenhagen, Denmark
\and
 Instituto de Astrof\'\i sica de Andaluc\'\i a (C.S.I.C.) Apartado 3004, 18080 Granada, Spain 
\and 
 Specola Vaticana, 00120 Stato Citt\`a del Vaticano}

\date{Received ...; accepted ...}

\titlerunning{Stellar populations in \wings \ galaxies}
\authorrunning{Fritz J. et al.}

\abstract 
{The WIde-field Nearby Galaxy clusters Survey (\wings) is a project whose primary goal is to study the galaxy populations in clusters in the local universe ($z<0.07$) and of the influence of environment on their stellar populations. This survey has provided the astronomical community with a high quality set of photometric and spectroscopic data for $77$ and $48$ nearby galaxy clusters, respectively.}
{In this paper we present the catalog containing the properties of galaxies observed by the \wings \ SPEctroscopic ({\sc wings-spe}) survey, which were derived using stellar populations synthesis modelling approach. We also check the consistency of our results with other data in the literature.}
{Using a spectrophotometric model that reproduces the main features of observed spectra by summing the theoretical spectra of simple stellar populations of different ages, we derive the stellar masses, star formation histories, average age and dust attenuation of galaxies in our sample.}
{$\sim 5300$ spectra were analyzed with spectrophotometric techniques, and this allowed us to derive the star formation history, stellar masses and ages, and extinction for the \wings \ spectroscopic sample that we present in this paper.}
{The comparison with the total mass values of the same galaxies derived by other authors based on \sdss \ data, confirms the reliability of the adopted methods and data.}

\keywords{methods: data analysis -- galaxies: clusters: general -- galaxies: fundamental parameters}

\maketitle


\section{INTRODUCTION}

One of the best places to study the influence of dense
environments on galaxy evolution are galaxy clusters. The fact
that early-type galaxies are more common in clusters, while spirals
are preferentially found in the field, is a manifestation of the so-called
morphology-density relation, which was discovered to be a
common pattern over a wide range of environmental densities, from local
groups of galaxies to distant clusters \citep[see, for example,
][]{postman84,dressler97,postman05}. Not only the morphology, but also
the stellar content of galaxies is influenced by the galaxy environment,
and clusters host galaxies with the oldest stellar populations.
Dense environments are capable of altering the star formation
history of a galaxy, quenching its star formation activity as it falls
 to the cluster, as a results of phenomena
such as gas stripping, tidal interactions, and/or gas starvation. 

While studies of the stellar populations are already available for distant
clusters \citep[see, e.g.,][]{poggianti99,poggianti08}, a similar
analysis on a homogeneous and complete set of data at low redshift
has been lacking until
now. \wings \ was conceived as a survey to serve as a local comparison for the
distant clusters studies. Thanks to its deep and
high-quality optical imaging and its large sample of cluster galaxy
spectra, it enables us to study in detail
the link between galaxy morphology and star formation
history.

Optical spectra
are nowadays widely exploited to derive the properties of the stellar
population content of galaxies, by means of spectral synthesis
techniques. In this paper we present the results of the
spectrophotometric analysis performed on the spectra of a sample of
local clusters galaxies from the \wings \ survey, describing how
stellar masses, star formation histories, dust attenuation and average
age are obtained.

 The \wings\footnote[1]{Please refer to the \wings \ website for
updated details about both the survey and its products;
\tt{http://web.oapd.inaf.it/wings/new/index.html}} project
\citep[see][]{fasano06} is providing the largest set of homogeneous
spectroscopic data for galaxies belonging to nearby
clusters. Originally designed as a B and V band photometric survey,
\wings \ has widened its database to also include near-infrared
bands \citep[J and K, see][]{valentinuzzi09} and ultraviolet
photometry (Omizzolo et al., in prep.), H$\alpha$ imaging (Vilchez et al., in
preparation) and optical spectroscopy \citep{cava09}. With such a
wealth of data, \wings \ has a considerable legacy value for
the astronomical community, becoming the local benchmark 
with which the properties of galaxies in high redshift clusters 
can be compared with.

In this paper, we present the
catalogs that we are providing as on-line databases and give a
full description of all the measurements and stellar population
properties that are given. In order to do so, we will summarize
the main features of our spectrophotometric model that are used to
derive such quantities, already described
in detail in previous work \citep[][F07 hereafter]{fritz07}.
Furthermore, in order to make all the
potential users of the databases more confident with the quantities
that we derive, we present a detailed and careful validation of
our results, by comparing the values obtained on a subsample
of \wings \ galaxies that are in common with the Sloan Digital Sky
Survey.

The paper outline is as follows: after
describing the \wings \ spectroscopic dataset in {\bf \S 2}, in {\bf \S 3}
we give a brief review of the adopted spectrophotometric model and recall the
characteristics of the theoretical spectra that are used; in {\bf \S 4} we describe
the properties of the stellar populations that are derived and how
they are computed, while in {\bf \S 5} we present a validation of our
results by comparing them with other literature data and, finally, in
{\bf \S 6}, we describe the items that will be provided in
the final catalog, and give an example.

We remind that the \wings \ project assumes a standard
 $\Lambda$ cold dark matter  ($\Lambda$CDM) cosmology with 
 H$_0=70$, $\Omega_\Lambda=0.70$ and $\Omega_M=0.30$.

\section{\wings \ SPECTROSCOPIC DATASET}\label{sec:data}
Out of the 77 cluster fields imaged by the \wings \ photometric survey
\citep{varela09}, 48 were also observed spectroscopically. While the
reader should refer to \cite{cava09} for a complete
description of the spectroscopic sample, (including completeness
analysis and quality check), here we will briefly summarize the
features that are more relevant for this work's purposes.\\ Medium
resolution spectra for $\sim 6000$ galaxies were obtained during
several runs at the 4.2m William Herschel Telescope (WHT) and at the 3.9m Anglo
Australian Telescope (AAT) with multifiber spectrographs (WYFFOS and
2dF, respectively), yielding reliable redshift measurements. The
fiber apertures were $1.6''$ and $2''$, respectively, and the spectral resolution 
$\sim 6$ and $\sim 9$ \AA \ FWHM for the WHT and AAT spectra, respectively. The
wavelength coverage ranges from $\sim 3590$ to $\sim 6800$ \AA \ for the
WHT observations, while spectra taken at the AAT covered the $\sim 3600$
to $\sim 8000$ \AA \ domain. Note also that, for just one observing run at the
WHT (in which 3 clusters were observed), the spectral resolution was
$\sim 3$ \AA \ FWHM, with the spectral coverage ranging from $\sim 3600$
to $\sim 6890$ \AA.

\section{THE METHOD}
We derive stellar masses, star formation histories, extinction values
and average stellar ages of galaxies by analysing their integrated
spectra by means of spectral synthesis techniques. The model that is
used for this analysis has already been described in detail in F07,
but here we will briefly and schematically recall its main features
and parameters.
\subsection{The fitting technique}\label{sec:fit}
The model reproduces the most important features of an observed
spectrum with a theoretical one, which is obtained by summing the
spectra of Single Stellar Population (SSP, hereafter) models of
different stellar ages and a fixed, common value of the metallicity. Before
being added together, each SSP spectrum is weighted with a proper value of
the stellar mass and dust extinction by an amount which,
in general, depends on the SSP age itself.

The best fit model parameters are obtained by calculating the
differences between the observed and model spectra, and evaluating
them by means of a standard $\chi^2$ function:
\begin{equation}\label{eqn:chi2}
\chi^2=\sum_{i=1}^{N} \left(\frac{M_i-O_i}{\sigma_i}\right)^2
\end{equation}
where $M_i$ and $O_i$ denote the quantities measured from the
model and observed spectra, respectively (i.e. continuum fluxes and
equivalent widths of spectral lines), with $\sigma_i$ being the
observed uncertainties and $N$ being the total number of observed
constraints. The observed errors on the flux are computed by taking
into account the local spectral signal-to-noise ration, while uncertainties
on the equivalent widths are derived mainly from the measurement
method (see section 2.2 in F07, and Fritz et al. 2010b, in prep., for
further details).

The observed features that are used to compare the likelihood between
the model and the observed spectra are chosen from the most
significant emission and absorption lines and continuum flux intervals. In
particular, we compare, when measurable. the equivalent widths of
H$\alpha$, H$\beta$, H$\delta$, H$\epsilon$+Ca{\sc ii (h)}, Ca{\sc ii
(k)}, H$\eta$ and {\sc [Oii]}. Other lines, even though prominent,
were only measured but not used to constrain the model parameters.
Key examples are the {\sc [Oiii]} line at 5007 \AA, because it is too sensitive to the
physical conditions of the gas and of the ionizing source, and the Na
and Mg lines at $\sim$5890 and 5177 \AA \ respectively, because they are
strongly affected by the enhancement of $\alpha$-elements, which is
not taken into account by our SSPs). The continuum flux is measured in
specific wavelength ranges, chosen to avoid any
important spectral line, so to sample as best as possible the
shape of the spectral continuum. 
Particular emphasis is given to the 4000 \AA \
break, D4000, defined by \cite{bruzual83}, as it is considered a good
indicator of the stellar age.

As already mentioned, the amount of dust extinction is let free
to vary as a function of the SSP age. Treating extinction in this way
is equivalent, in some sense, to taking into account the fact that
the youngest stars are expected to be still embedded within the dusty
molecular clouds where they formed, while as they become older,
they progressively emerge from them. In this picture, the spectra of
SSPs of different ages are supposed to be dust-reddened by
different amounts; dust is assumed to be distributed so to simulate a uniform
layer in front of the stars, and the Galactic extinction curve \citep{cardelli89}
is adopted.

Building a self-consistent chemical model, that would take into
account changes in the metal content of a galaxy and its chemical
evolution as a function of mass and star formation history, was far
beyond the scope of this work. This is why we adopted a homogeneous
value for the metallicity for our theoretical spectra, and left it to the model
free to choose between three
different sets of metallicity, namely Z=0.05, Z=0.02 and Z=0.004
(super-solar, solar and sub-solar, respectively). Fitting an observed
spectrum with a single value of the metallicity is equivalent to
assuming that this value belongs to the stellar population that is
dominating its light. However, as described in F07, acceptable fits
are obtained for most of the spectra adopting different metallicities,
which means that this kind of analysis is often not able to
provide a unique value for the metallicity.

It is clear that, assuming a unique value for the SSP's metallicity
when reproducing an observed spectrum is a simplifying
assumption since, in practice, the stellar populations of a galaxy
span a range in metallicity values. One could hence
question the reliability of the mass and of the SFH determination
done by using one single metallicity value. To better understand 
this possible bias due to the mix of metallicities that is expected
in galaxies, we repeated the check already performed in F07: we built
template synthetic spectra with 26 different SFHs as in F07, but with
values of the metallicity varying as a function of stellar age, to 
roughly simulate a chemical evolution, and we analyzed them by
means of our spectrophotometric fitting code. The results clearly show
that the way we deal with the metallicity does not introduce any bias
in the recovered total stellar mass or SFH.

\subsection{SSP parameters}\label{sec:ssp}
All of the stellar population properties that are derived are
strictly related to the theoretical models that we use in our fitting
algorithm. It is hence of foundamental importance to give all the
details of the physics and of the parameters that were used to build
them.

First of all, WE make use of the Padova evolutionary tracks
\citep{bertelli94} and use a standard \cite{salpeter55} initial mass
function (IMF), with masses in the range 0.15-120 M$_\odot$. Our
optical spectra were obtained using two different sets of observed
stellar atmospheres: for ages younger than $10^9$ years we used
\cite{jacoby84}, while for older SSPs we used spectra from the MILES
library \citep{Psanchez04,Psanchez06} and both sets were degraded in
spectral resolution, in order to match that of our observed spectra
(namely 3, 6 and 9 \AA \ of FWHM, see Sect.\ref{sec:data} for
details). Using the theoretical libraries of Kurucz, the SSP spectra were
extended to the ultra-violet and infrared, widening, in this way, the
wavelength range down to 90 and up to $\sim 10^9$ \AA \ (note that in
these intervals the spectral resolution is much lower, being $\sim 20$
\AA, but in any case outside the range of interest for the spectra
used for our analysis).

Gas emission, whose effect is visible through emission lines, was also 
computed --and included
in the theoretical spectra-- by means of the photoionisation code {\sc
cloudy} \citep{ferland96}. The optical spectra of SSPs younger than
$\sim 2\times 10^7$ display, in this way, both permitted and forbidden
lines (typically, hydrogen, {\sc [Oii]}, {\sc [Oiii]}, {\sc [Nii]} and
{\sc [Sii]}). This nebular component was computed assuming {\it case B
recombination} \citep[see][]{osterbrock89}, an electron temperature
of $10^4$ K, and an electron density of $100$ cm$^{-3}$. The radius of the
ionizing star cluster was assumed to be 15 pc, and its mass  $10^4$
M$_\odot$. Finally, emission from the circumstellar envelopes of AGB
stars was computed and added as described in \cite{bressan98}.

The initial set of SSPs was composed of 108 theoretical spectra referring
to stellar ages ranging from $10^5$ to $20\times 10^9$ years, for each one
of the three afore-mentioned values of the metallicity. Determining 
the age of stellar populations
from an integrated optical spectrum with such a high temporal
resolution is well beyond the capabilities of any spectral
analysis. Hence, as a first step, we reduced the stellar age
resolution by binning the spectra. This was done by taking into
account both the characteristics of the evolutionary phases of stars,
and the trends in spectral features as a function of the SSP age (see both
section 2.1.1 and Fig. 1 in F07). After combining the spectra at this
first stage, we ended up with 13 stellar age bins.

\begin{figure*}[t]
\centering
\begin{tabular}{rr}
\rotatebox{270}{
\includegraphics[width=0.37\textwidth]{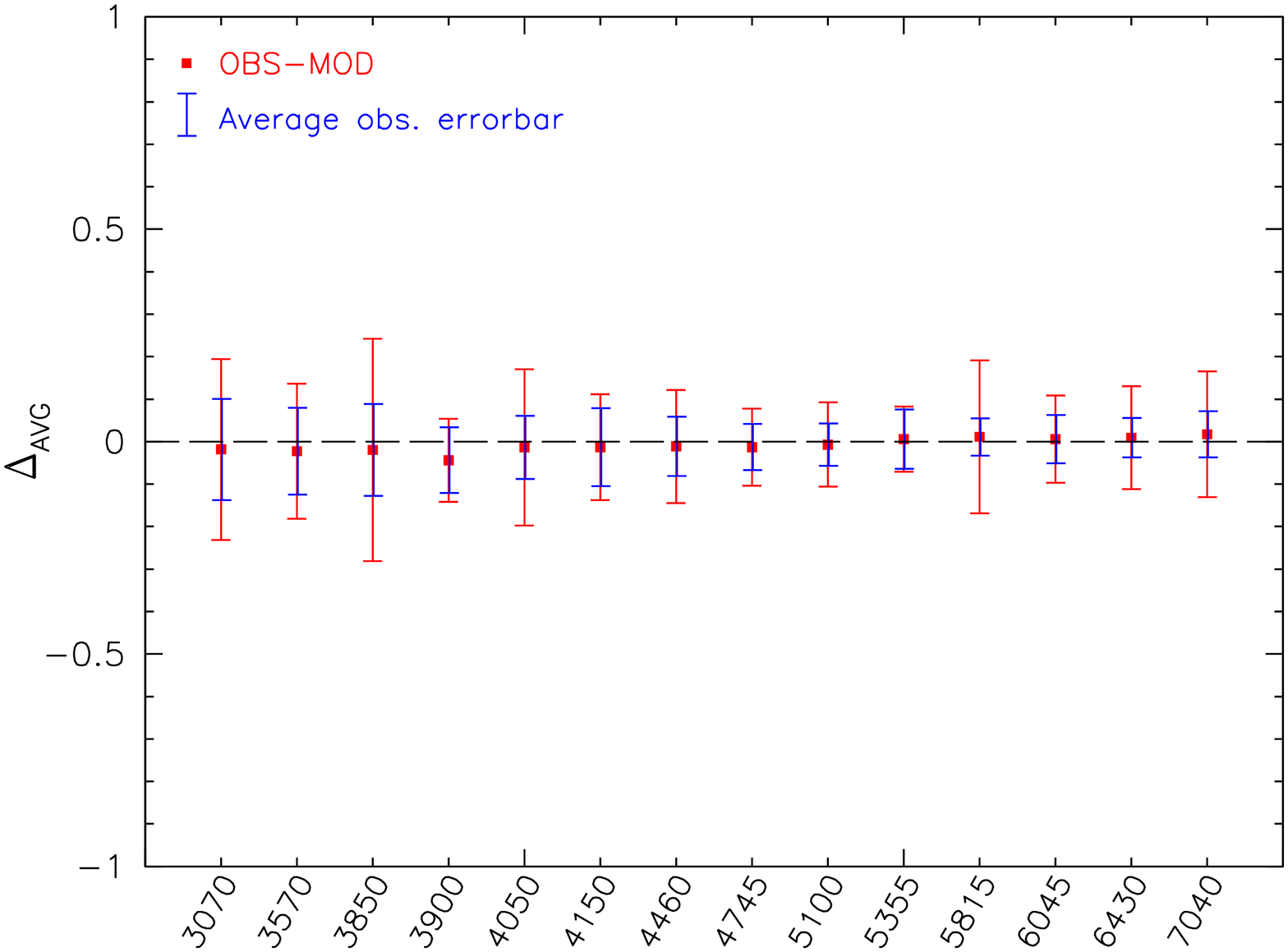}}&
\rotatebox{270}{
\includegraphics[width=0.37\textwidth]{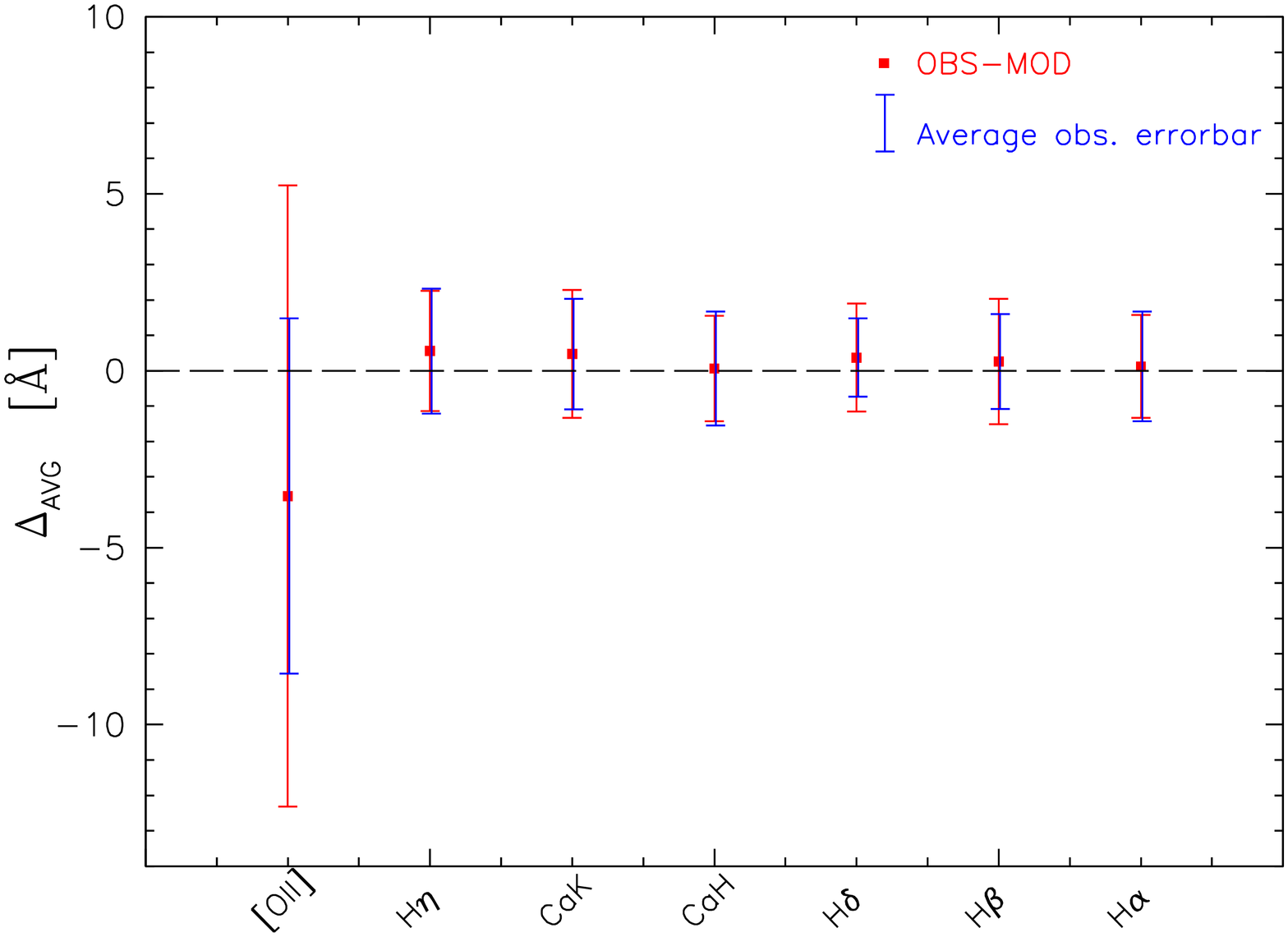}}\\
\end{tabular}
\caption{In this figures we plot the values of the difference, $\Delta$, 
between spectral features in the observed and model spectra, averaged 
over all the spectra of the \wings \ sample with an acceptable spectral fit
($\chi^2<3$). On the left panel we show 
the differences calculated for the continuum flux (both observed and model
spectra have been normalized to 1 at 5500 \AA), 
and on the right panel for the equivalent width of the 
lines. Red points represent the average value of $\Delta$, 
for each one of the continuum bands and emission lines that were used as 
constraints in the fit. The red errorbars are the corresponding rms, 
while blue ones are the average of the observed rms.}
\label{fig:rms}
\end{figure*}
As we describe in F07, this set of theoretical spectra originally
included also a SSP whose age, namely $\sim 17.5$ Gyr, is older than
the universe age. The use of this SSP was merely statistical: since
the only appreciable difference between the three oldest SSPs of our
set is, actually, the mass-to-luminosity ratio, using such an old SSP
would prevent our random search of the best fit model to be
systematically biased towards the youngest of the old
SSPs. Nevertheless, the adoption of such an approach can lead some models
to be dominated by this very old stellar population yielding, in this
way, mass values that are too high, due to the higher mass-to-light
ratio. To overcome this issue we decide to avoid the use of the oldest
stellar populations, limiting ourselves to stellar populations whose
ages are consistent with that of the universe. We will hence refer,
from now on, to these 12 SSPs.

\subsection{The best fit search}
Finding the best combination of the parameters that minimizes the
differences between the observed and the model spectrum, is a
non-linear problem, due to the presence of extinction. Furthermore, it
is also underdetermined, which means that the number of constraints is
lower than the number of parameters. In fact, in our case, we are
using SSPs of 12 different ages, so that our task turns into finding
the combination of 12 mass and extinction values that better fits the
observed spectrum. To find the set of 24 parameters that will yield
the best fit model, we use the Adaptive Simulated Annealing algorithm,
which randomly explores the parameters space, searching for an absolute
minimum in the $\chi^2$ function. This method is particulary suited to
such problems, where the function to minimize has lots of local
minima: once a promising zone for a minimum, in the parameter space,
is found, the algorithm not only refines the search of the local
minimum, but also checks for the presence of other, deeper minima,
outside the local ``low-$\chi^2$ valley''.

\subsection{Uncertainties}
All the physical parameters that are derived from the the spectral analysis,
refer to a best fit
model for an observed spectrum. The limited wavelength range under
analysis, the well known age-metallicity degeneracy, and the non-linearity
of the problem, together with the fact that it is underdetermined,
makes the solution non-unique. This means that models with different
characteristics may equally well reproduce the observed spectral
features. To account for this, we give error-bars related to mass,
extinction and age values.

To compute such uncertainties, we exploit the characteristics of the
minimisation algorithm: the path towards the best fit model (or the minimum
$\chi^2$) depends on
the starting points so, in general, starting from different initial positions can
lead to different minimum points, i.e. to best fit models with
different parameters. We hence perform 11 optimisations, each time
starting from a different point in the parameters space. In this way we
end up with 11 best fit models that we verified are well
representative of the space of the solutions. We take, as a reference,
the model with the median total mass among these 11. All the errorbars
are computed as the average difference between the values of the
models with the highest and lowest total stellar mass.

\subsection{The quality of the fits}
The similarity between an observed spectrum and its best fit model is
measured, as explained in \S \ \ref{sec:fit}, by means of a $\chi^2$ function
taking into account both spectral continuum fluxes and the equivalent 
widths of significant lines. Our choice to use a wide range both
in metallicity and SSP ages, and to let both extinction and mass vary freely,
are the key ingredients that allow us to satisfactorily reproduce
any galactic spectrum, at least in principle. 

In practice, low quality spectra
due to low S/N, bad flux calibration, bad subtraction of sky or telluric
lines, can give rise to a bad fit. To demonstrate that
there are no systematic failures of any of the observed features that
are used as constraints, in Fig.\ref{fig:rms} we show the difference
between the values calculated for the model and
for the observed spectrum, averaged over all the \wings \ sample. In
the left-hand panel we show, plotted as red squares, the average 
values of the difference for the flux in the spectral continuum,
together with the rms (red errorbars), and the average values of the
observed errors (blue errorbars). 

The plot in the right-hand panel of the same figure shows the differences
for the equivalent widths of the spectral lines.  The {\sc [Oii]} line
is the one that shows the highest displacement with respect to the
zero-difference line, due to the fact that this line is in the
spectral region with the highest noise. This makes it also more
difficult to measure, and it also explains why its observed value has
the average largest error. 
Overall all the features are well reproduced, with no systematic failure.

\section{THE PROPERTIES OF STELLAR POPULATIONS}
In this section we describe the properties of the stellar populations
that are derived from our spectrophotometric synthesis, that are now
publicly available. Fitting the main features of an
optical spectrum allows us to derive the characteristics of the stellar
populations whose light we see in the integrated spectrum: total mass,
mass of stars as a function of age, the metallicity and dust
extinction are typical quantities that can be obtained. As already
pointed out, using this particular technique, it is almost
impossible to recover a unique value for the stellar metallicity due to both
the degeneracy issues such as the age-metallicity and
age-extinction and to the fact that we do not consider SSP
models with $\alpha$-element enhancements. In fact, in the vast
majority of cases at least two values of the metallicity are found to
provide equally good fits.

\subsection{Stellar masses}\label{sec:mass}
When stellar masses are derived by means of spectrophotometric
techniques, it is important to clearly state which definition of mass
is used. As already made clear by Longhetti \& Saracco \citep[2009, but see
also][]{renzini06}, the use of spectral synthesis techniques leads to
three different definitions of the stellar mass, namely:
\begin{enumerate}
\item the initial mass of the SSP, at age zero; this is nothing but
the mass of gas turned into stars
\item the mass locked into stars, both those which are still in the
nuclear-burning phase, and remnants such as white dwarfs, neutron
stars and stellar black holes
\item the mass of stars that are still shining, i.e. in a nuclear-burning phase. 
\end{enumerate}
The difference between the three definitions is a function of the stellar 
age and, in particular, it can be up to a factor of 2 between mass definition 
1) and 3), in the oldest stellar populations. We will provide the user 
with masses calculated using all of the afore mentioned definitions, 
following the same enumeration.

\begin{figure}
\includegraphics[height=0.5\textwidth]{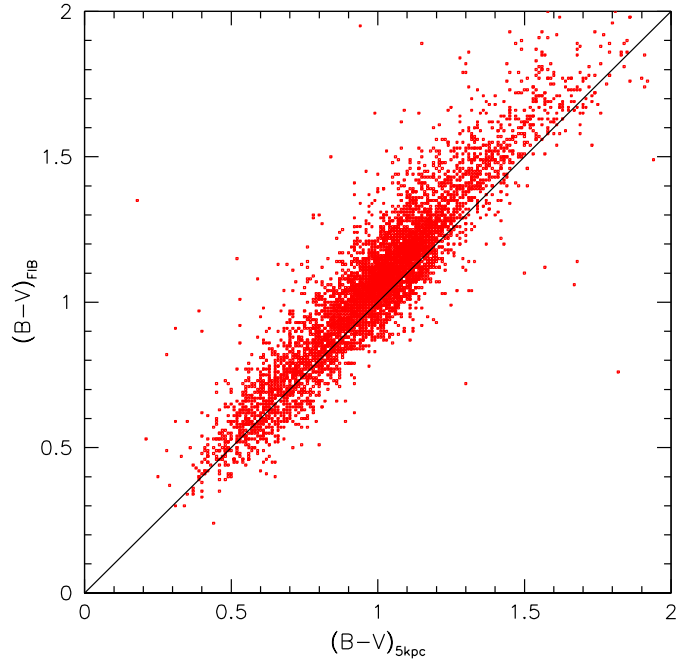}
\caption{The comparison between values of the $(B-V)$ colour as computed 
from a 5 kpc aperture (x axis) 
and the fiber aperture (y axis) magnitudes. The solid line represents the 1:1 relation 
that highlights a systematic, off-set of $\sim 0.1$ mag: the 5 kpc colour is bluer as 
expected since the total magnitude is sampling, on average, younger populations in the 
outskirts of the galaxies.}
\label{fig:colorgrad}
\end{figure}

To compute the values of stellar mass, we exploit the fact that the
theoretical spectra are given in luminosity per unit of solar
mass. Once the model spectrum is converted to flux by accounting for
the luminosity distance factor, the K-correction is naturally
performed by fitting the spectra at their observed redshifts. All of
the observed spectra are normalised by means of their observed V-band
magnitude within the fiber aperture.
Obviously, in order to obtain a stellar mass value referring to
the whole galaxy (that we will dub ``total stellar mass'', from now
on), one should use a spectrum representative of the whole galaxy,
which is not at our disposal. Since we have both aperture and total
photometry for all the objects of our spectroscopic sample, we use the
total V magnitude to rescale the model spectrum: in this way we are
assuming that the colour gradient of the aperture-to-total magnitude
is negligible (this assumption is made by several authors: see
e.g. \citealt{kauffmann03}). When speaking of ``total magnitude''
here, we refer to the \texttt{MAG\_AUTO} value \citep[see][for further
details]{varela09}, that is the \texttt{SExtractor} magnitude computed
within the Kron aperture.\\ In Fig. \ref{fig:colorgrad} we show the
comparison between the observed $(B-V)$ colour computed using the
magnitudes within a 5 kpc aperture (x axis) an the one computed using 
the spectroscopic
fiber aperture (y axis), with the black line being the 1:1
relation. We consider the color within 5 kpc a good approximation
of the total color (in fact, it closely follows the color derived using
AUTO magnitudes), and is a good aperture compromise for both
large and small galaxies in our sample.  
The average difference between the fibre and 5 kpc colours is $\sim 0.1$
mag, due to the presence of bluer (thus probably younger) stars in the
outskirts of the galaxies. We will provide values of the
stellar mass referring to both apertures, and a colour term which can
be used to correct the total mass to account for radial gradients in
the stellar populations content, as described below.

\subsection{Colour corrections}\label{sec:colcor}

To correct total masses for colour gradients,
we exploit the \cite{bdj01} prescription, which was derived in order
to compute stellar masses in galaxies by means of photometric data. 
According to their work, the M/L ratio of a
galaxy can be expressed by the following:
\begin{equation}\label{eqn:bdj}
\log_{10}\left( \frac{M}{L_\lambda}\right)=a_\lambda+b_\lambda\cdot COL
\end{equation}
where $L_\lambda$ is the luminosity in a given band (denoted by
$\lambda$) while the $a_\lambda$ and $b_\lambda$ coefficients depend
on the band that is used, and on the population synthesis models
(including IMF, isochrones, etc.), and $COL$ is the colour term. Table
4 in \cite{bdj01} presents a list of such coefficients for various
bands, models and two metallicities (subsolar ---Z=0.008--- and solar
---Z=0.02---). For the calculations that follow, we will use V and B
band data, and assume the \cite{kodama97} models, that use a Salpeter
IMF and a solar metallicity value, which yields $a_V=-0.18$ and
$b_V=1.00$. Note that, using \cite{bruzual03} or {\sc pegase}
\citep{fioc97} models, will not substantially affect the results.

As already mentioned above, when going from the stellar mass
calculated over the fiber magnitude to the one referring to the whole
galaxy, there is the implicit assumption that the colour calculated
within the fiber aperture is the same as the one calculated with the
total magnitudes (here we assume a $\sim 5$ kpc aperture), while this
is not true for most cases. Starting from Eq.\ref{eqn:bdj} and after
some algebra, we derive a colour-correction term as follows:
\begin{equation}\label{eqn:ccol}
C_{corr}= b_V\cdot [(B_5-V_5)-(B_f-V_f)]
\end{equation}
where the term $(B_5-V_5)$ is the colour computed from 5 kpc aperture
magnitudes and $(B_f-V_f)$ is the observed colour within the
fiber. 
This factor, which is given in our final catalogs, must be
added to the total mass value in order to account for colour
gradients.

As a consistency check for the values of the total stellar mass
computed by means of our models, we compare them to the values that
can be obtained by means of Eq.\ref{eqn:bdj}, which yields the
following:
\begin{equation}\label{eqn:bdj2}
\log_{10}\frac{M}{M_\odot}=-0.4\cdot(V-V_\odot)+a_V+b_V\cdot (B^k_5-V^k_5)
\end{equation}
where $B^k_5-V^k_5$ are the K-corrected (i.e. rest-frame) magnitudes
extracted from a 5 kpc aperture, $V$ is the total absolute
magnitude (obtained from V \texttt{MAG\_AUTO}, K-corrected)
and $V_\odot=4.82$ is the absolute magnitude of the sun in
the V band. K-corrections were taken from \cite{poggianti97}.  

\begin{figure} 
\centering
\includegraphics[height=.5\textwidth]{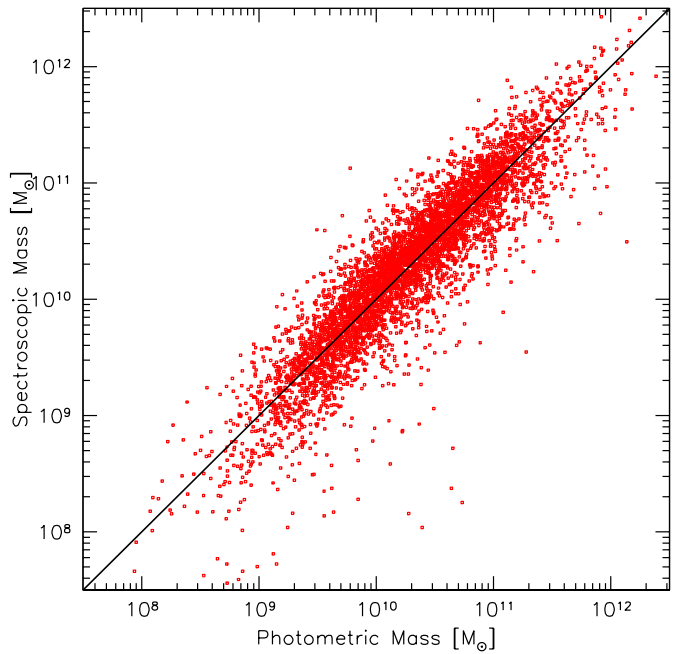}
\caption{A comparison of the total stellar mass of galaxies in the
\wings \ sample, as computed by means of B and V band
photometry, on the $x$-axis, assuming the prescriptions given in
\cite{bdj01} (see text for details), and by means of our spectral
fitting. The solid line represents the 1:1 relation.
A colour correction
term, computed as explained in the text, was applied to the
spectroscopic-derived values, while the photometric values were
corrected to account for the difference in the IMF mass limits.}
\label{fig:bdj}
\end{figure}

In Fig.\ref{fig:bdj} we show the comparison between total stellar
masses computed by means of our spectral fitting (on the $y$-axis) and
those obtained by means of the \cite{bdj01} prescription (i.e. by adopting
Eq.\ref{eqn:bdj2}). We applied a 0.064 dex correction to account for
the differences in the adopted IMF (\citealt{bdj01} use a Salpeter
IMF with masses in the 0.1--100 M$_\odot$ range, while we use
0.15--120 M$_\odot$), and we added the
colour correction term to the spectroscopic-derived mass values, as
explained above. The agreement between the two different methods is,
on average, always better than 0.1 dex. A similar comparison between
stellar masses obtained from spectral fitting and from photometry, calculated
using aperture magnitudes instead of the total ones, shows an equally good
agreement between the two methods.
The \cite{bdj01} mass photometric values are also
provided in our final catalogs.

\subsection{The star formation history}\label{sec:sfh}
As we describe in F07 and summarize in Sect. \ref{sec:ssp}, our search
for the best fit-model is performed using 12 SSPs of different ages,
obtained, in turn, by binning a much higher age-resolution stellar age
grid. Still, we verified that it is not possible to recover the star
formation as a function of stellar age with the relatively high temporal
resolution provided by the 12 SSPs. 
After performing accurate tests on template spectra that
were built in order to match the spectral features of \wings \ spectra
in terms of both spectral resolution, signal-to-noise ratio and
wavelength coverage, we found that it is possible to properly recover
the star formation history (hereafter, SFH) 
in 4 main stellar age bins. The details of the choice are
explained in F07; here we just recall their ranges that are,
respectively: $0-2\times 10^7$, $2\times 10^7-6\times 10^8$,
$6\times 10^8-5.6\times 10^9$ and $5.6\times 10^9-14\times
10^9$ years.

The SFH is given in our catalogs in two different forms: 
1) percentage of the stellar
mass and 2) star formation rate (SFR) in the four bins. The first is
computed according to the following:
\begin{equation}\label{eqn:massbin}
M_{bin}=\sum_{i=1}^{N_{bin}}\left(C_i\times M^\star_i \right) / \sum_{i=1}^{N_{SSP}}\left(C_i\times M^\star_i \right) 
\end{equation}
where $N_{bin}$ is the number of SSPs contained in a given age bin; $C
_i$ is the normalisation constant of each SSP of that bin, i.e. the
stellar mass at each age according to definition 1; $M^\star_i$ is the
factor, which is a function of the stellar age, that converts the SSP
initial mass (definition 1.) into either the mass locked into stars
(mass definition 2.) or into mass of nuclear burning stars
(definition 3.), while the sum at the denominator is the total
stellar mass (according to definitions 2 and 3, respectively).

The star formation rate as a function of the stellar age is computed
by dividing the stellar mass of a given age bin by its
duration. Definition 1 of the mass was applied in this calculation
\citep[see also equation 1 in][]{longhetti09}.

The current SFR value, i.e. the one calculated within the
youngest age bin, deserves a particular attention, since it is
calculated by fitting the equivalent width of emission lines, namely
Hydrogen (H$\alpha$ and H$\beta$) and Oxygen ({\sc [Oii]} at 3727
\AA).  The lines' luminosity is entirely attributed to
star formation processes neglecting other mechanisms
that can produce ionizing flux. In this way we are overestimating
the current SFR in both LINERS and AGNs. 
In a forthcoming work, we will present an analysis of standard
diagnostic diagrams such as those by \cite{veilleux87}, with the
lines' intensities accurately measured by subtracting stellar 
templates from the observed spectrum (Marziani et al., in prep.).
This work will enable
the distinction between ``pure'' star forming systems and those where
other mechanisms might be co-responsible for line emission.

\subsection{Dust extinction}
According to the ``selective extinction'' hypothesis
\citep{calzetti94}, which we fully consider in our modelling, each SSP
has its own value of the dust attenuation. We compute an age-averaged
value of dust extinction, as it is derived by the model, by using
Eq.\ref{eqn:av}:
\begin{equation}\label{eqn:av}
A_V=-2.5\times \log_{10}\left[\frac{L_{tot}^M(5550)}{L_{unext}^M(5550)} \right]
\end{equation}
where $L_{tot}^M$ and $L_{unext}^M$ are, respectively, the model
spectrum and the model non-attenuated spectrum (i.e. the model with
the same SFH as $L_{tot}^M$ but with $A_V=0$ for each stellar
population). 
We calculate two distinct values: we first take into account only stellar 
populations that are younger than $\sim 2\times 10^7$,
i.e. those that are responsible for nebular emission; this value is comparable
with extinction that is computed from emission lines ratio. Secondly, 
we use all stellar populations providing, in this way, an extinction value 
which is averaged over SSP of all ages.

\subsection{Average ages}
Exploiting the information derived by our analysis,
we are able to provide an estimate of the average age of a galaxy,
weighted on the stellar populations that compose its spectrum. Given
that the mass-to-light ratio changes as a function of the age, there
are two different definitions that can be given: the mass-weighted and
the luminosity-weighted age \citep[see also][]{fernandes03}. The
latter is the most commonly given, since it is directly derived from
the spectrum, being weighted in this way towards the age of the stellar
populations that dominate the light, while the first definition
requires the knowledge of the mass distribution as a function of
stellar age, i.e. the SFH. 
We can compute the logarithm of these two quantities as follows:
\begin{equation}\label{eqn:lwage}
\langle \log(T)\rangle_L=\frac{1}{L_{tot} (V)}\times \sum_{i=1}^{N_{SSP}}L_i(V)\times \log(t_i)
\end{equation}
for the logarithm of the luminosity weighted age, where $L_i(V)$  and $L_{tot}(V)$ are the restframe luminosities of the {\it i-th} SSP and of the total spectrum, respectively, in the V-band, and $t_i$ the age of the $i-th$ SSP. The mass-weighted age is computed in a similar way as:
\begin{equation}\label{eqn:mwage}
\langle \log(T)\rangle_M=\frac{1}{M_{tot}}\times \sum_{i=1}^{N_{SSP}}M_i\times \log(t_i)
\end{equation}
and, similarly, $M_{tot}$ and $M_i$ are the total mass and the mass of the {\it i-th} SSP, respectively. Hence, while the luminosity-weighted age gives an estimate of the age of stars that dominate the optical spectrum, being in this way more sensitive to the presence of young stars, the mass-weighted value is more representative of the actual average age of a galaxy's stellar populations. Note that to compute these values, we use the finest age grid, averaging over the 12 stellar populations.

We provide both the luminosity-weighted age computed from the V-band, and the one computed from the bolometric luminosity. The two values are, anyway, very similar.

\subsection{Absolute magnitude computation and prediction}\label{sec:mag}
The fact that the theoretical SSP spectra that we use for our modeling
cover a wide range in wavelengths, allows us to compute absolute
magnitudes in various bands that are not covered by the observed
spectra, without having to assume any K-correction. To compute the
absolute magnitude of a galaxy, we take the best-fit model spectrum,
compute its flux as if it was observed at 10 pc and convolve it with
the proper filter transmission curve:
\begin{equation}
M_b=\frac{\int_{\lambda_0}^{\lambda_1} F_{d=10pc}^M(\lambda)\times T_b(\lambda) \; d\lambda}{\int_{\lambda_0}^{\lambda_1} T_b(\lambda) \; d\lambda}
\end{equation}
where $T_b(\lambda)$ is the transmission curve of the filter for the band $b$ and $F_{d=10pc}^M(\lambda)$ is the model spectrum calculated at a distance of 10 pc. For the sake of clearity, in Table \ref{tab:f0} we provide the zero-point fluxes, expressed in erg/s/cm$^2$/\AA \ that were used to compute all of the magnitudes.
\begin{table}
\centering
\begin{tabular}{ccc}
      $f_0$        & $\lambda_{eff}$ & Band \\
\hline 
[erg/s/cm$^2$/\AA] &     [\AA]       &      \\
\hline
\hline
4.217e-09 	   &	  3605       &  U   \\
6.600e-09 	   &	  4413       &  B   \\
3.440e-09 	   &	  5512       &  V   \\
1.749e-09 	   &	  6586       &  R   \\
8.396e-10 	   &	  8060       &  I   \\
3.076e-10 	   &	 12370       &  J   \\
1.259e-10 	   &	 16460       &  H   \\
4.000e-11 	   &	 22100       &  K   \\
8.604e-09 	   &	  3521       &  u   \\
4.676e-09 	   &	  4804       &  g   \\
2.777e-09 	   &	  6253       &  r   \\
1.849e-09 	   &	  7668       &  i   \\
1.315e-09 	   &	  9115       &  z   \\
\hline
\end{tabular}
\caption{Zero-point fluxes that are used to calculate observed expected 
magnitudes and absolute magnitudes, together with their effective lambda. 
Johnson and \sdss \  filters characteristics were taken from the Asiago 
Database of Photometric Systems \citep{moro00}.}
\label{tab:f0}
\end{table}
$UBVRIJHK$ magnitudes are computed according to the Johnson system, while 
$ugriz$ magnitudes are calculated in order to match the Sloan system.

\section{VALIDATION}
In order to compare with the widely used \sdss \ masses,
we performed a comparison of the stellar mass values for a subsample of
\wings \ galaxies that has been spectroscopically observed also by the
\sdss. As a reference for masses from the SLOAN survey we used those
derived by \cite{gallazzi05}, using the Data Release 4
(DR4)\footnote{Stellar masses computed by \cite{gallazzi05}, by means
of DR4 data are publicly available at this website:
\tiny{\texttt{http://www.mpa-garching.mpg.de/SDSS/DR4/Data/stellarmet.html}}},
and those obtained from the photometry exploiting Data Release 7
(DR7)\footnote{Stellar mass values for the DR7 data release were taken
from the following \sdss \ website:
\tiny{\texttt{http://www.mpa-garching.mpg.de/SDSS/DR7/Data/stellarmass.html}}}. 
In this way, we built two sub-samples of galaxies observed by both surveys,
namely 395 in the \wings-DR4 sample, and 606 in the \wings-DR7 sample.
\begin{figure*} [!t]
\centering
\begin{tabular}{ll}
\includegraphics[height=.495\textwidth]{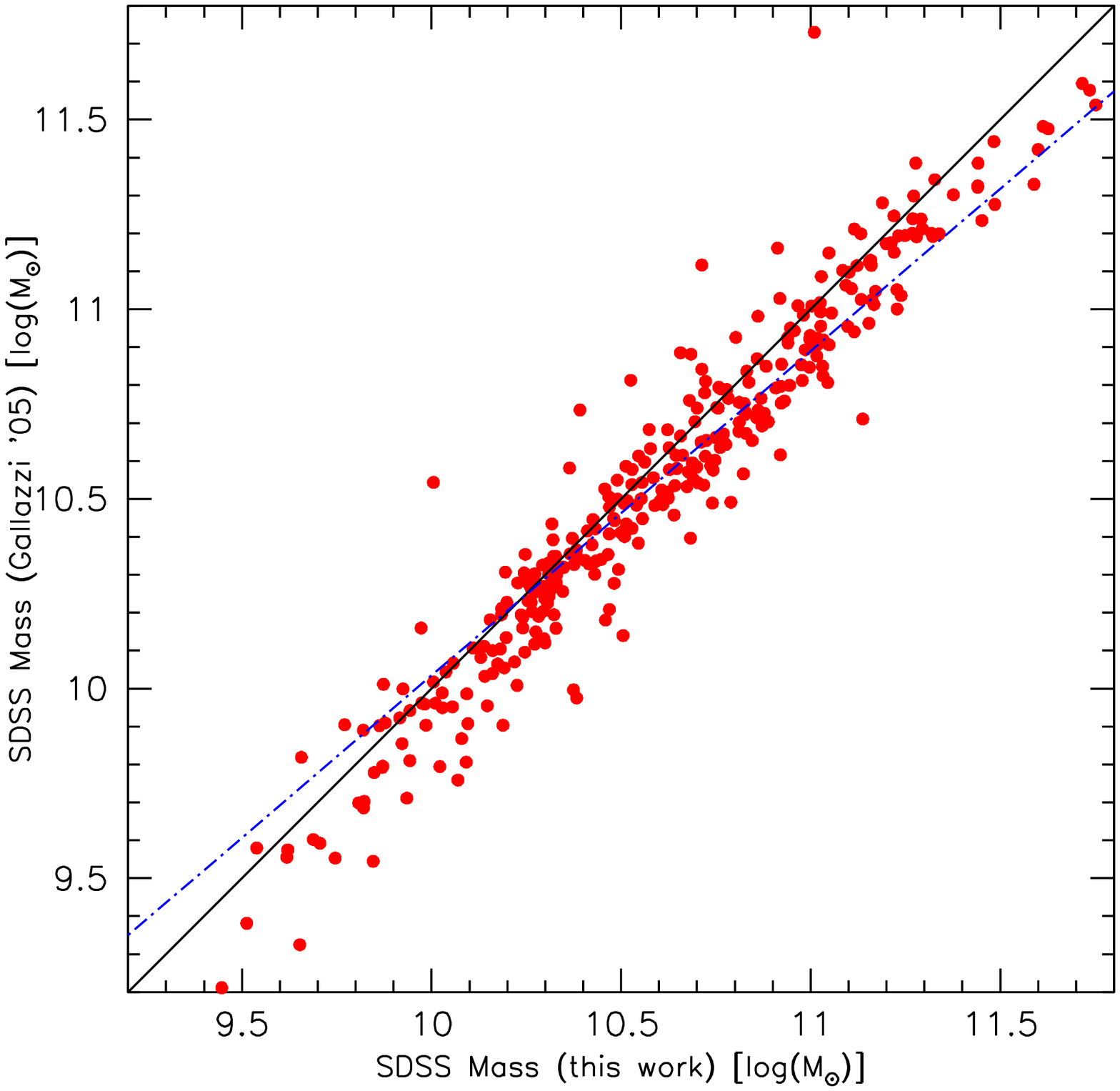} &
\includegraphics[height=.495\textwidth]{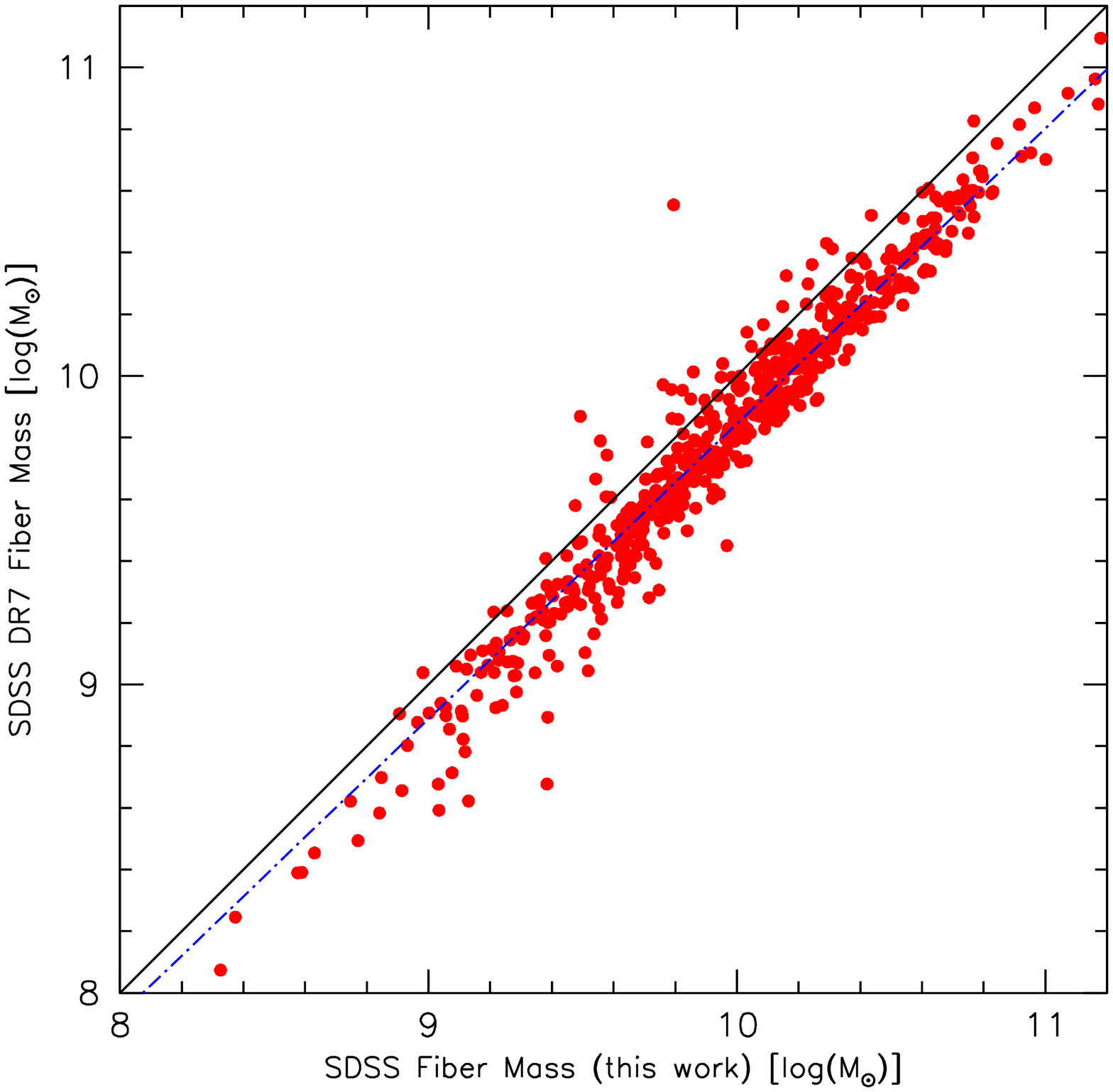} \\
\end{tabular}
\caption{In the left panel we show the comparison between mass values
that we obtained by fitting \sdss \ spectra with our model, and
those calculated by \cite{gallazzi05}. The black line represents the
1:1 relation and the blue dotted-dashed line is the least-square fit
to the data. On the right, we compare mass values we derived using our
spectrophotometric fitting on the sloan's fiber spectrum, to those
obtained from DR7 photometric data fitting, referring to the same
aperture. Lines and symbols as in the left panel. All sets of mass
values have been corrected to account for differences in the assumed
IMF (see text for details), and those that are shown here are
normalized to the \cite{kroupa01} IMF.}
\label{fig:mass_comp}
\end{figure*}

We performed a double check: as a first step, we exploited our
spectrophotometric model to derive, using \sdss \ spectra and $g$
(model-)band magnitudes of the \wings-DR4 sample, the same
quantities that were inferred for \wings \ galaxies. In this way,
comparing the results obtained with the same (\sdss) 
data but with different methods,
we can demonstrate the reliability of our technique. As a second step, we
compare total stellar mass values obtained with our model and \wings \
data, to those of the \sdss \ DR4 and DR7, respectively.

To ensure this comparison is significant, we have to consider the
details of the models used to derive such quantities. In particular,
we have to take into account the differences in the IMFs that are
assumed, i.e.  \cite{salpeter55} for \wings \ (we recall here that the
mass limits that we have adopted are 0.15 and 120 M$_\odot$,
respectively), \cite{chabrier03} for masses derived by
\cite{gallazzi05}, and \cite{kroupa01} for \sdss, DR7, respectively. We
have determined that the difference between Salpeter's and
Kroupa's IMF is a factor of $\sim1.33$ (0.125 dex), the Salpeter IMF
yielding the highest values of masses, while Chabrier's IMF yields
stellar masses that are 1.1 (0.04 dex) times lower with respect to 
Kroupa's \citep[see, e.g.,][]{cimatti08}. For the sake of homogeneity,
and only for the purposes of these sanity checks, we will rescale all
the mass values to the \cite{kroupa01} IMF. Note that all the mass
values we refer to are calculated according to definition 2. (see
Sect.\ref{sec:mass}).

In Fig.\ref{fig:mass_comp} we show how the different methods
compare, exploiting both DR4 and DR7 data. In the left-hand panel, 
we plot mass values derived using our
model against those obtained by \cite{gallazzi05}, both from \sdss \ DR4
data. Our mass determination was obtained by fitting the \sdss \
spectrum ---which was normalized to the total model $g$ band
magnitude--- in the same way as done for \wings \ data. The agreement
between the two methods overall is good, with an rms of $\sim 0.21$.

\begin{figure} 
\centering
\begin{tabular}{ll}
\includegraphics[height=.48\textwidth]{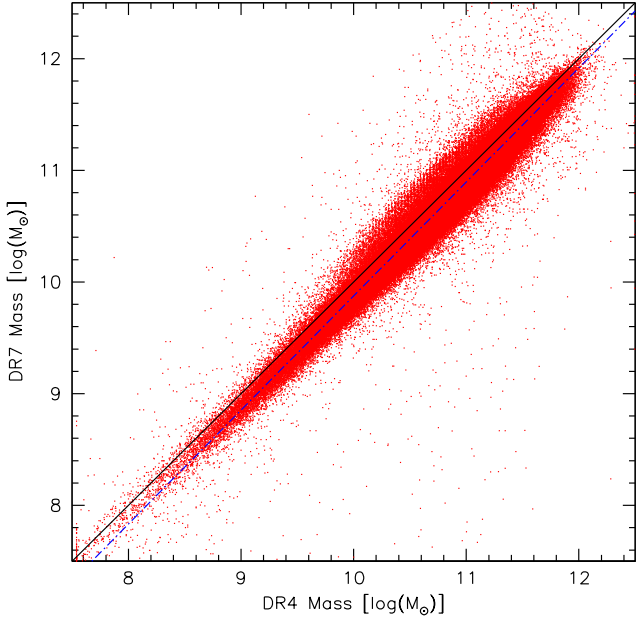} 
\end{tabular}
\caption{The comparison between total mass values
from the \sdss \ DR4, as calculated by \cite{gallazzi05}, and those
obtained from DR7 photometry. Both masses are rescaled to a 
Kroupa (2001) IMF.}
\label{fig:sdss2}
\end{figure}

In the right-hand panel of Fig.\ref{fig:mass_comp} we show the comparison
between the masses we derived from the \sdss \ spectroscopy 
scaled to the $g$ band fiber magnitude
and the fiber-aperture photometric masses from the DR7. 
Hence, we are comparing two mass estimates within the same
fibre aperture, obtained using either the \sdss \ spectroscopy+photometry
or only \sdss \ photometry, and we do not have to deal with aperture effects.
The rms is $\sim 0.17$, but it is worth noting that,
the data displays a $\sim 0.15$dex systematic offset,
in the sense that the DR7 yields slightly lower masses. 
This is in contrast to the DR4 comparison which shows remarkable agreement, 
even though there is some dispersion with respect to the 1:1 relation.
A small offset in the same direction
is present also when comparing DR4 and DR7 masses for galaxies
in common, as shown in Fig.\ref{fig:sdss2}.

\begin{figure*} 
\centering
\begin{tabular}{ll}
\includegraphics[height=.48\textwidth]{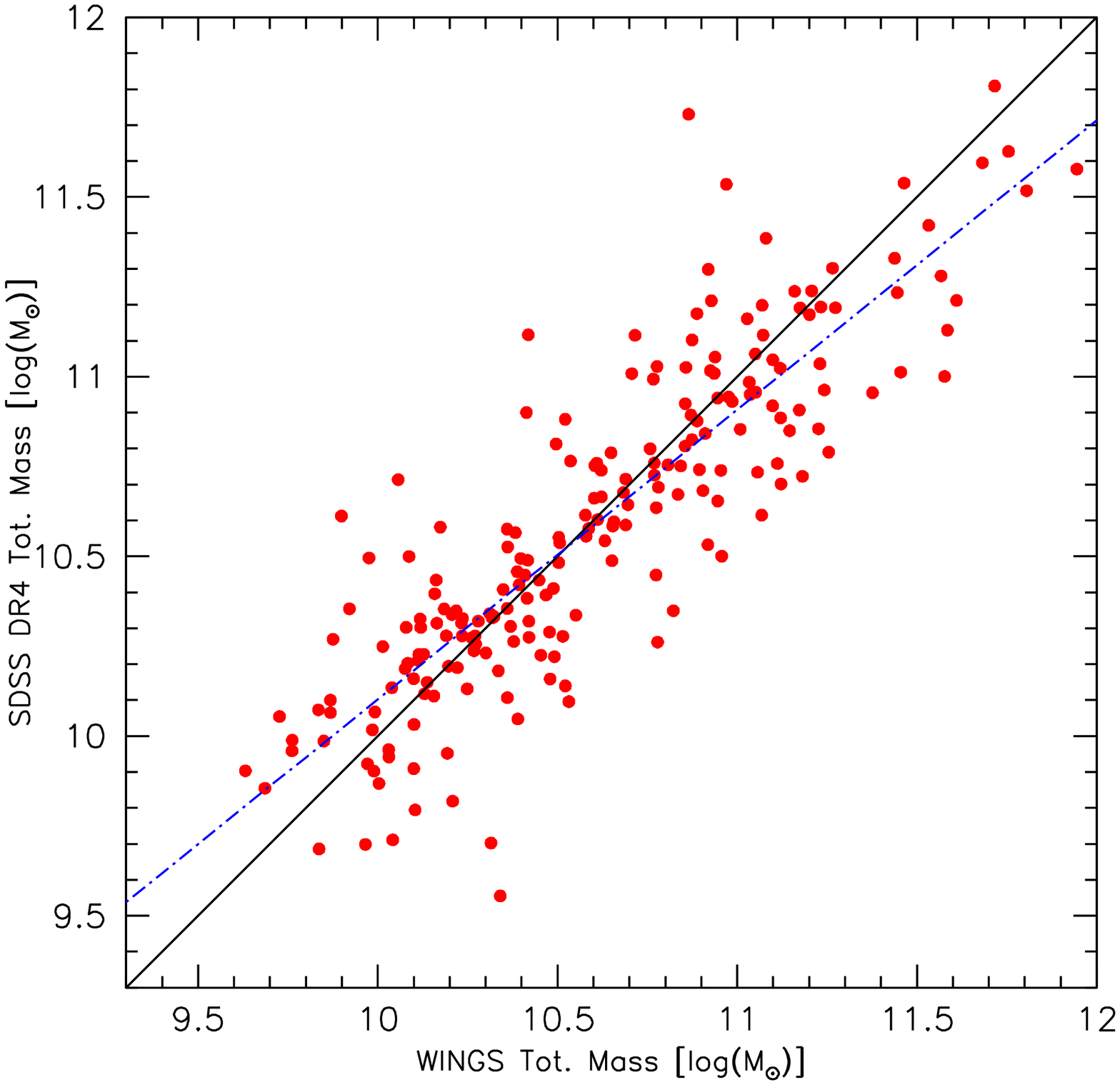} &
\includegraphics[height=.48\textwidth]{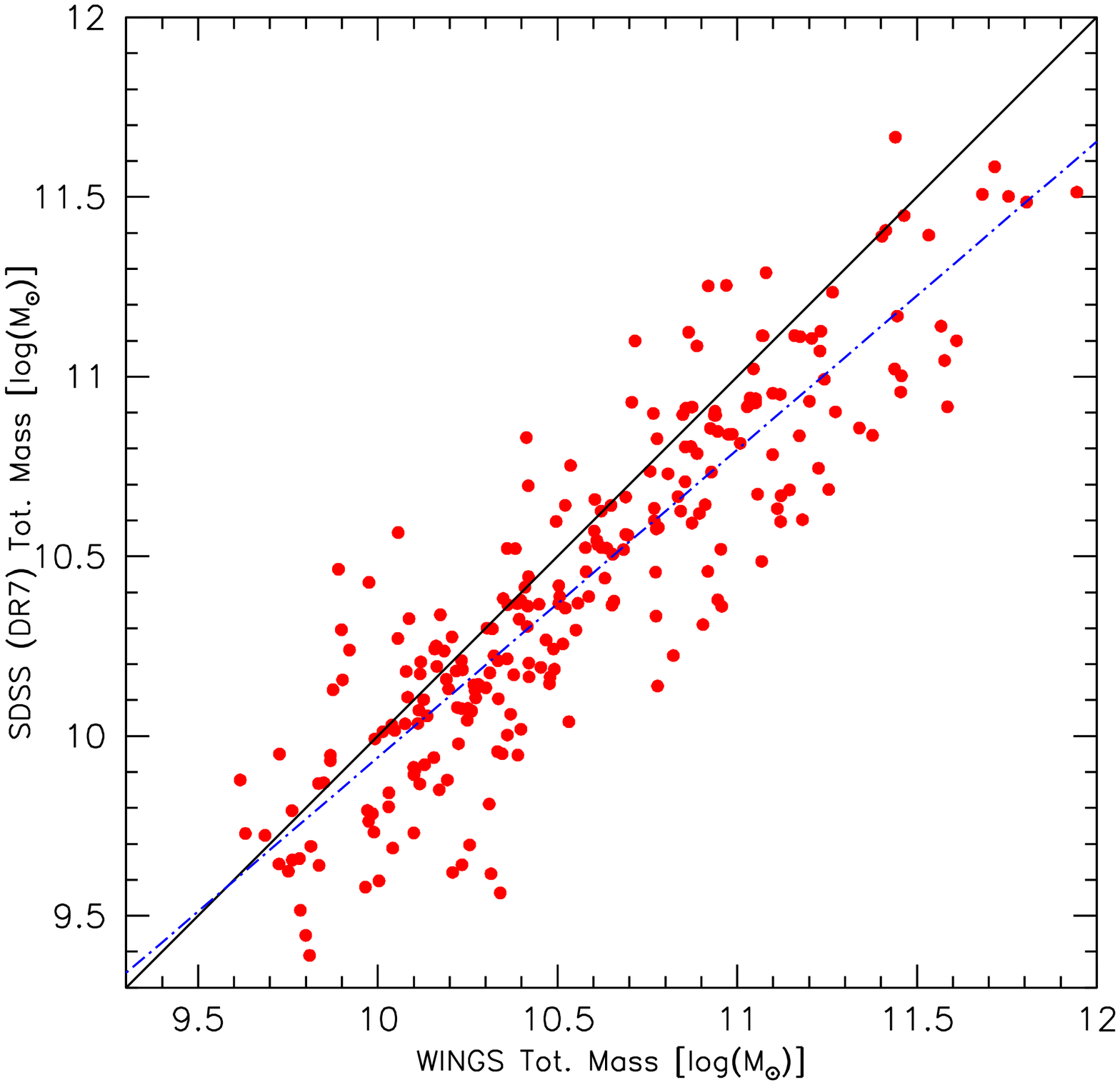} \\
\end{tabular}
\caption{In the left panel we show the comparison between mass values
from the \sdss \ DR4, as calculated by \cite{gallazzi05}, and those
obtained by means of our spectrophotometric model on \wings \ spectra
and magnitudes. On the right panel we present the same comparison, but
with DR7 masses, derived from total magnitude. The large scatter is
due to the combination of both different methods and different
data. The black line is the 1:1 relation, and the red line is the
least-square fit to the data for both plots. All the values are
rescaled to the \cite{kroupa01} IMF, and a colour correction term has
also been applied to \wings \ masses (see Sect.\ref{sec:colcor}).}
\label{fig:mass_comp2}
\end{figure*}

In Fig.\ref{fig:mass_comp2} we move to the comparison of the total
stellar masses we derived from the \wings \ data, corrected for color
gradients, and total masses given by Gallazzi et al. (2004)
and DR7,
always considering galaxies of the subsamples in common.
The scatter around the 1:1 relation is slightly larger in these cases,
(0.23 and 0.22 for DR4 and DR7, respectively)
and for the DR7-derived masses the average
difference is negligible at low masses and
tends to increase with mass.
For this comparison, in addition to the different mass estimate methods,
the data are also different: \wings \ spectra are
taken within an aperture of $\sim 2''$ while Sloan fibers cover a
$\sim 3''$ aperture, centered on a position that can be, in general,
different. 
Also, source and flux extraction techniques, and the spectral
resolution of the two surveys are different.
The general agreement, however, is satisfactory, and the scatter is similar
to the $\sim 0.2$dex accuracy expected with these methods (e.g.
Cimatti et al. 2008).

We conclude that, despite the substantial differences in the fitting
approach, in the adopted theoretical libraries and in the
characteristics of the datasets themselves, our total mass values are
in overall agreement with those of a considerable number of objects
that the \sdss \ has in common with \wings.

\section{THE CATALOGS}
In this section we briefly describe the most relevant quantities given
in the catalogs we are releasing to the
astronomical community. About 70\% of the observed spectra have been
fitted with a $\chi^2 \leq 3$ and we consider fits with such values 
to be reliable. For higher values, a visual inspection is recommended 
to asses the reliability of the spectral fit. 

For each spectrum that has been analyzed we give the following:
\begin{itemize}
\item the reduced $\chi^2$ for the fits obtained for the three
values of metallicity. Note that we take as a
reference model the one with the value of the metallicity that yields
the lowest $\chi^2$ value, regardless of the fact that other values of
the metallicity are also providing acceptable fits. These values are
also useful to flag potentially unreliable fits. A $\chi^2 \leq 3$ can be used as
a discriminant for blindly accepting a result;
\item extinction in the V band, in magnitudes, computed from the model
spectrum both averaging on young stellar populations (i.e. with
age$\leq 2\times 10^7$ years) and on all ages, including uncertainties
on both quantities;
\item SFR in the four main age bins as defined in Sect.\ref{sec:sfh},
with related uncertainties, all expressed in $M_\odot/yr$; note that
these SFR only refer to values normalized to the fiber-aperture
magnitude. In order to compute the global value, one should multiply
the fiber-SFR by a factor $\wp=10^{-0.4\cdot (V_{tot}-V_{fib})}$, that is
the ratio of total and aperture fluxes;
\item percentage of the stellar mass in the 4 main age bins, with
related uncertainties, calculated for the different mass definitions
\item the logarithm of total stellar mass, expressed in $M_\odot$,
within the fiber aperture, according to the 3 definitions explained in
section \ref{sec:mass}, together with the related uncertainties
(expressed in logarithm of solar masses as well), which are computed
for the definition 3. mass value
\item the logarithm of total stellar mass, expressed in $M_\odot$,
computed by rescaling the fiber spectrum to the total V magnitude (see
section \ref{sec:mass}), and the related uncertainties (in logarithm of
the solar mass), uncorrected for color gradients;
\item the logarithm of the stellar mass calculated from the B and V
band photometry, according to the
\cite{bdj01} prescription, for both total and fiber magnitudes;
\item the colour-correction term, described in \S4.2, 
to be added to the total mass to account for color gradients;
\item the logarithm of the luminosity-weighted age computed 
 both using the luminosity in the V band, and the bolometric
emission, and the related uncertainties: the latter are computed only
with respect to the bolometric luminosity-weighted age;
\item the logarithm of the mass-weighted age, and the related uncertainty
\item Galactic extinction-corrected observed B and V magnitude
referring to both the fiber and the total aperture; we report these
magnitudes even though they are actually measured values
\citep[see][]{varela09}, because these are the values used to rescale
the observed spectrum and, hence, to derive the total mass. Values of
extinction within our Galaxy for each of the clusters were taken from NED
\citep[see also][]{schlegel98};
\item absolute V and B magnitudes calculated from the observed
spectrum, derived from both aperture and total magnitudes;
\item Johnson (UBVIRJHK) and Sloan (ugriz) expected observed
magnitudes calculated from our best model spectrum, both within our
fiber aperture and total;
\item Johnson (UBVIRJHK) and Sloan (ugriz) absolute magnitudes
calculated from our best model spectrum, both within our fiber
aperture and total.
\end{itemize}
Whenever one of the above listed quantities is not available, this is
flagged with a 99.99.

All the data and physical quantities described in this paper will be
available by querying the \wings \ database at the following web
address:\\
\texttt{http://web.oapd.inaf.it/wings/}. 

In Table \ref{tab:sfh} we 
give an example of how the full set of information will look like, 
reporting data for 5 galaxies of the sample. A description of each
item, together with their units, can be found in table \ref{tab:header}.\\

\noindent{}{}
{\bf ACKNOWLEDGEMENTS}\\

\noindent{}{}This paper took great advantage from discussions with 
Anna Gallazzi and Jarle Brinchmann, who kindly provided us with 
all the details of their stellar masses calculations using DR4 and DR7, 
respectively. \\ 
Funding for the SDSS and SDSS-II has been provided by the Alfred 
P. Sloan Foundation, the Participating Institutions, the National Science 
Foundation, the U.S. Department of Energy, the National Aeronautics 
and Space Administration, the Japanese Monbukagakusho, the Max 
Planck Society, and the Higher Education Funding Council for England.\\
The SDSS Web Site is: \texttt{http://www.sdss.org/}. \\
We are grateful to the anonymous referee, whose comments and remarks
helped us to improve the quality and the readability of this work.


\clearpage
\onecolumn
\begin{landscape}
\begin{table}
\scriptsize
\caption{ }
\vspace{0.5cm}
\label{tab:sfh}
\begin{tabular}{|l||l|r|r|r|r|r|r|r|r|r|r|r|r|r|}
\hline
  \multicolumn{1}{|c||}{WINGSID} &
  \multicolumn{1}{c|}{SPENAME} &
  \multicolumn{1}{c|}{L\_DIST} &
  \multicolumn{1}{c|}{$\chi^2_{Z05}$} &
  \multicolumn{1}{c|}{$\chi^2_{Z02}$} &
  \multicolumn{1}{c|}{$\chi^2_{Z004}$} &
  \multicolumn{1}{c|}{Z} &
  \multicolumn{1}{c|}{AV\_Y} &
  \multicolumn{1}{c|}{AV\_Y\_E} &
  \multicolumn{1}{c|}{AV\_T} &
  \multicolumn{1}{c|}{AV\_T\_E} &
  \multicolumn{1}{c|}{SFR1} &
  \multicolumn{1}{c|}{SFR1\_E} &
  \multicolumn{1}{c|}{SFR2} &
  \multicolumn{1}{c|}{SFR2\_E} \\
\hline 
WINGSJ103833.76-085623.3 & A1069\_12\_129 & 234.64  & 7.193  & 7.108  & 7.215  & 0.020 & 0.426 & 0.504 & 0.241 & 0.184 & 0.0527 & 0.0377 & 0.0328 & 0.0226 \\
WINGSJ103834.09-085719.2 & A1069\_12\_144 & 247.74  & 11.378 & 11.296 & 11.214 & 0.004 & 0.931 & 0.252 & 0.342 & 0.107 & 0.0058 & 0.0190 & 0.0320 & 0.0068 \\
WINGSJ103835.89-085031.5 & A1069\_11\_163 & 856.69  & 1.568  & 1.624  & 9.674  & 0.050 & 1.045 & 0.320 & 0.200 & 0.059 & 0.3143 & 0.2649 & 0.1988 & 0.0918 \\
WINGSJ103842.69-084611.6 & A1069\_11\_132 & 908.99  & 9.000  & 8.351  & 8.407  & 0.020 & 0.621 & 0.240 & 0.165 & 0.079 & 0.3780 & 0.1027 & 1.1693 & 0.1998 \\
WINGSJ103843.03-085602.8 & A1069\_11\_171 & 250.29  & 1.548  & 1.395  & 1.832  & 0.020 & 1.299 & 0.241 & 0.400 & 0.134 & 1.7552 & 1.2031 & 0.0721 & 0.0566 \\
\hline
\end{tabular}
\\[5pt]
\begin{tabular}{|l||r|r|r|r|r|r|r|r|r|r|r|r|r|r|r|r|r|r|r|r|r|}
\hline
  \multicolumn{1}{|c||}{ } &
  \multicolumn{1}{c|}{SFR3} &
  \multicolumn{1}{c|}{SFR3\_E} &
  \multicolumn{1}{c|}{SFR4} &
  \multicolumn{1}{c|}{SFR4\_E} &
  \multicolumn{1}{c|}{M1\_1} &
  \multicolumn{1}{c|}{M1\_2} &
  \multicolumn{1}{c|}{M1\_3} &
  \multicolumn{1}{c|}{M1\_E} &
  \multicolumn{1}{c|}{M2\_1} &
  \multicolumn{1}{c|}{M2\_2} &
  \multicolumn{1}{c|}{M2\_3} &
  \multicolumn{1}{c|}{M2\_E} &
  \multicolumn{1}{c|}{M3\_1} &
  \multicolumn{1}{c|}{M3\_2} &
  \multicolumn{1}{c|}{M3\_3} &
  \multicolumn{1}{c|}{M3\_E} &
  \multicolumn{1}{c|}{M4\_1} &
  \multicolumn{1}{c|}{M4\_2} \\
\hline 
... & 0.0320 & 0.0134 & 0.0073 & 0.0156 & 0.0044& 0.0054 & 0.0059 & 0.0050 & 0.0787 & 0.0877 & 0.0941 & 0.0774 & 0.6617 & 0.6560 & 0.6559 & 0.2682 & 0.2553 & 0.2509 \\
... & 0.0014 & 0.0028 & 0.0080 & 0.0047 & 0.0012& 0.0015 & 0.0018 & 0.0073 & 0.1979 & 0.2167 & 0.2397 & 0.0900 & 0.0745 & 0.0732 & 0.0754 & 0.1497 & 0.7263 & 0.7086 \\
... & 4.2464 & 3.6629 & 8.1438 & 6.6400 & 0.0001& 0.0001 & 0.0001 & 0.0001 & 0.0013 & 0.0014 & 0.0015 & 0.0006 & 0.2344 & 0.2343 & 0.2408 & 0.3196 & 0.7642 & 0.7643 \\
... & 0.0828 & 0.0992 & 0.4542 & 0.3301 & 0.0015& 0.0020 & 0.0022 & 0.0008 & 0.1367 & 0.1462 & 0.1573 & 0.0737 & 0.0835 & 0.0847 & 0.0889 & 0.0995 & 0.7783 & 0.7672 \\
... & 0.6712 & 0.4775 & 1.3432 & 0.6312 & 0.0024& 0.0030 & 0.0034 & 0.0026 & 0.0028 & 0.0032 & 0.0036 & 0.0031 & 0.2260 & 0.2264 & 0.2318 & 0.1758 & 0.7688 & 0.7674 \\
\hline
\end{tabular}
\\[5pt]
\begin{tabular}{|l||r|r|r|r|r|r|r|r|r|r|r|r|r|r|r|}
\hline
  \multicolumn{1}{|c||}{ } &
  \multicolumn{1}{c|}{M4\_3} &
  \multicolumn{1}{c|}{M4\_E} &
  \multicolumn{1}{c|}{M\_1\_FIB} &
  \multicolumn{1}{c|}{M\_2\_FIB} &
  \multicolumn{1}{c|}{M\_3\_FIB} &
  \multicolumn{1}{c|}{M\_FIB\_E}   &
  \multicolumn{1}{c|}{M\_1\_TOT} &
  \multicolumn{1}{c|}{M\_2\_TOT} &
  \multicolumn{1}{c|}{M\_3\_TOT} &
  \multicolumn{1}{c|}{M\_TOT\_E} &
  \multicolumn{1}{c|}{AM\_BJ} &
  \multicolumn{1}{c|}{TM\_BJ} &
  \multicolumn{1}{c|}{CCOL} &
  \multicolumn{1}{c|}{L\_AGE\_V} &
  \multicolumn{1}{c|}{L\_AGE} \\
\hline
...  & 0.2440 & 0.2206 & 8.3839  & 8.2619  & 8.2169  & 7.7948  & 9.0159  & 8.8939  & 8.8489  & 8.4267  & 8.4184  & 9.0504  & 0.1210  & 8.41915 & 8.14517 \\
...  & 0.6832 & 0.1753 & 7.9725  & 7.8784  & 7.8131  & 7.2090  & 8.6765  & 8.5824  & 8.5171  & 7.9129  & 8.2291  & 8.9330  & -0.0390 & 8.34588 & 8.32951 \\
...  & 0.7576 & 0.3193 & 10.9570 & 10.8295 & 10.7690 & 10.0297 & 11.4170 & 11.2895 & 11.2290 & 10.4895 & 10.8701 & 11.3301 & -0.1890 & 9.86304 & 9.88039 \\
...  & 0.7516 & 0.1461 & 9.6955  & 9.5724  & 9.5191  & 8.9408  & 10.0235 & 9.9004  & 9.8471  & 9.2687  & 9.6927  & 10.0207 & -0.0190 & 8.63437 & 8.27387 \\
...  & 0.7612 & 0.1768 & 10.1717 & 10.0432 & 9.9849  & 9.1670  & 10.6597 & 10.5312 & 10.4729 & 9.6546  & 10.0673 & 10.5553 & 0.0610  & 9.31117 & 9.29801 \\
\hline
\end{tabular}
\\[5pt]
\begin{tabular}{|l||r|r|r|r|r|r|r|r|r|r|r|r|}
\hline
  \multicolumn{1}{|c||}{} &
  \multicolumn{1}{c|}{L\_AGE\_E} &
  \multicolumn{1}{c|}{M\_AGE} &
  \multicolumn{1}{c|}{M\_AGE\_E} &
  \multicolumn{1}{c|}{V\_PH\_FIB} &
  \multicolumn{1}{c|}{B\_PH\_FIB} &
  \multicolumn{1}{c|}{V\_PH\_TOT} &
  \multicolumn{1}{c|}{B\_PH\_TOT} &
  \multicolumn{1}{c|}{V\_OB\_FIB} &
  \multicolumn{1}{c|}{B\_OB\_FIB} &
  \multicolumn{1}{c|}{V\_OB\_TOT} &
  \multicolumn{1}{c|}{B\_OB\_TOT} &
  \multicolumn{1}{c|}{U\_MOD\_FIB} \\
\hline
...  & 7.65407 & 9.62516  & 9.33765 & 20.984 & 21.353 & 19.404 & 19.833 & -15.845 & -15.432 & -17.426 & -17.013 & 21.040 \\ 
...  & 7.62478 & 9.70064  & 9.22101 & 21.474 & 21.923 & 19.714 & 20.073 & -15.424 & -15.112 & -17.184 & -16.872 & 21.598 \\ 
...  & 9.08572 & 10.00135 & 9.37758 & 19.764 & 21.263 & 18.614 & 19.883 &  99.990 & -19.176 &  99.990 & -20.325 & 21.850 \\ 
...  & 7.43700 & 9.83851  & 9.36962 & 20.944 & 21.433 & 20.124 & 20.553 &  99.990 & -18.331 &  99.990 & -19.151 & 21.676 \\ 
...  & 8.71111 & 10.00002 & 9.24891 & 18.294 & 18.873 & 17.074 & 17.683 & -18.774 & -17.898 & -19.993 & -19.117 & 19.217 \\ 
\hline
\end{tabular}
\\[5pt]
\begin{tabular}{|l||r|r|r|r|r|r|r|r|r|r|r|r|}
\hline
  \multicolumn{1}{|c||}{ } &
  \multicolumn{1}{c|}{B\_MOD\_FIB} &
  \multicolumn{1}{c|}{V\_MOD\_FIB} &
  \multicolumn{1}{c|}{R\_MOD\_FIB} &
  \multicolumn{1}{c|}{I\_MOD\_FIB} &
  \multicolumn{1}{c|}{J\_MOD\_FIB} &
  \multicolumn{1}{c|}{H\_MOD\_FIB} &
  \multicolumn{1}{c|}{K\_MOD\_FIB} &
  \multicolumn{1}{c|}{U\_MOD\_T} &
  \multicolumn{1}{c|}{B\_MOD\_T} &
  \multicolumn{1}{c|}{V\_MOD\_T} &
  \multicolumn{1}{c|}{R\_MOD\_T} &
  \multicolumn{1}{c|}{I\_MOD\_T} \\
\hline
...  & 21.467 & 20.999 & 20.511 & 20.006 & 19.425 & 18.837 & 18.441 & 19.460 & 19.887 & 19.419 & 18.931 & 18.426 \\
...  & 21.922 & 21.442 & 21.050 & 20.630 & 20.277 & 19.829 & 19.480 & 19.838 & 20.162 & 19.682 & 19.290 & 18.870 \\
...  & 21.415 & 19.737 & 18.833 & 18.022 & 16.974 & 16.268 & 15.528 & 20.700 & 20.265 & 18.587 & 17.683 & 16.872 \\
...  & 21.995 & 21.033 & 20.436 & 19.950 & 19.511 & 19.004 & 18.332 & 20.856 & 21.175 & 20.213 & 19.616 & 19.130 \\
...  & 19.259 & 18.225 & 17.535 & 16.817 & 15.980 & 15.326 & 14.894 & 17.997 & 18.039 & 17.005 & 16.315 & 15.597 \\
\hline
\end{tabular}
\\[5pt]
\begin{tabular}{|l||r|r|r|r|r|r|r|r|r|r|r|r|r|r|}
\hline
  \multicolumn{1}{|c||}{ } &
  \multicolumn{1}{c|}{J\_MOD\_T} &
  \multicolumn{1}{c|}{H\_MOD\_T} &
  \multicolumn{1}{c|}{K\_MOD\_T} &
  \multicolumn{1}{c|}{U\_ABS\_F} &
  \multicolumn{1}{c|}{B\_ABS\_F} &
  \multicolumn{1}{c|}{V\_ABS\_F} &
  \multicolumn{1}{c|}{R\_ABS\_F} &
  \multicolumn{1}{c|}{I\_ABS\_F} &
  \multicolumn{1}{c|}{J\_ABS\_F} &
  \multicolumn{1}{c|}{H\_ABS\_F} &
  \multicolumn{1}{c|}{K\_ABS\_F} &
  \multicolumn{1}{c|}{U\_ABS\_T} &
  \multicolumn{1}{c|}{B\_ABS\_T} &
  \multicolumn{1}{c|}{V\_ABS\_T} \\
\hline
... & 17.845 & 17.257 & 16.861 & -15.870 & -15.416 & -15.823 & -16.329 & -16.841 & -17.408 & -17.991 & -18.261 & -17.450 & -16.996 & -17.403 \\
... & 18.517 & 18.069 & 17.720 & -15.463 & -15.094 & -15.458 & -15.889 & -16.307 & -16.650 & -17.085 & -17.329 & -17.223 & -16.854 & -17.218 \\
... & 15.824 & 15.118 & 14.378 & -18.647 & -19.199 & -20.276 & -21.043 & -21.818 & -22.763 & -23.449 & -23.753 & -19.797 & -20.349 & -21.426 \\
... & 18.691 & 18.184 & 17.512 & -18.420 & -18.212 & -18.730 & -19.237 & -19.650 & -20.201 & -20.725 & -21.017 & -19.240 & -19.032 & -19.550 \\
... & 14.760 & 14.106 & 13.674 & -17.910 & -17.921 & -18.790 & -19.509 & -20.222 & -21.013 & -21.653 & -21.943 & -19.130 & -19.141 & -20.010 \\
\hline
\end{tabular}
\\[5pt]
\begin{tabular}{|l||r|r|r|r|r|r|r|r|r|r|r|r|r|}
\hline
  \multicolumn{1}{|c||}{ } &
  \multicolumn{1}{c|}{R\_ABS\_T} &
  \multicolumn{1}{c|}{I\_ABS\_T} &
  \multicolumn{1}{c|}{J\_ABS\_T} &
  \multicolumn{1}{c|}{H\_ABS\_T} &
  \multicolumn{1}{c|}{K\_ABS\_T} &
  \multicolumn{1}{c|}{u\_MOD\_F} &
  \multicolumn{1}{c|}{g\_MOD\_F} &
  \multicolumn{1}{c|}{r\_MOD\_F} &
  \multicolumn{1}{c|}{i\_MOD\_F} &
  \multicolumn{1}{c|}{z\_MOD\_F} &
  \multicolumn{1}{c|}{u\_MOD\_T} &
  \multicolumn{1}{c|}{g\_MOD\_T} &
  \multicolumn{1}{c|}{r\_MOD\_T} \\
\hline
...  & -17.909 & -18.421 & -18.988 & -19.571 & -19.841 & 21.795 & 21.152 & 20.952 & 20.787 & 20.635 & 20.215 & 19.572 & 19.372 \\
...  & -17.649 & -18.067 & -18.410 & -18.845 & -19.089 & 22.355 & 21.584 & 21.478 & 21.384 & 21.317 & 20.595 & 19.824 & 19.718 \\
...  & -22.193 & -22.968 & -23.913 & -24.599 & -24.903 & 22.758 & 20.581 & 19.350 & 18.866 & 18.508 & 21.608 & 19.431 & 18.200 \\
...  & -20.057 & -20.470 & -21.021 & -21.545 & -21.837 & 22.454 & 21.508 & 20.884 & 20.749 & 20.748 & 21.634 & 20.688 & 20.064 \\
...  & -20.729 & -21.442 & -22.233 & -22.873 & -23.163 & 20.021 & 18.691 & 18.031 & 17.641 & 17.341 & 18.801 & 17.471 & 16.811 \\
\hline
\end{tabular}
\\[5pt]
\begin{tabular}{|l||r|r|r|r|r|r|r|r|r|r|r|r|}
\hline
  \multicolumn{1}{|c||}{ } &
  \multicolumn{1}{c|}{i\_MOD\_T} &
  \multicolumn{1}{c|}{z\_MOD\_T} &
  \multicolumn{1}{c|}{u\_ABS\_F} &
  \multicolumn{1}{c|}{g\_ABS\_F} &
  \multicolumn{1}{c|}{r\_ABS\_F} &
  \multicolumn{1}{c|}{i\_ABS\_F} &
  \multicolumn{1}{c|}{z\_ABS\_F} &
  \multicolumn{1}{c|}{u\_ABS\_T} &
  \multicolumn{1}{c|}{g\_ABS\_T} &
  \multicolumn{1}{c|}{r\_ABS\_T} &
  \multicolumn{1}{c|}{i\_ABS\_T} &
  \multicolumn{1}{c|}{z\_ABS\_T} \\
\hline
...  & 19.207 & 19.055 & -15.083 & -15.695 & -15.895 & -16.043 & -16.223 & -16.663 & -17.275 & -17.475 & -17.623 & -17.803 \\ 
...  & 19.624 & 19.557 & -14.658 & -15.371 & -15.471 & -15.534 & -15.638 & -16.418 & -17.131 & -17.231 & -17.294 & -17.398 \\ 
...  & 17.716 & 17.358 & -17.777 & -19.777 & -20.548 & -20.961 & -21.310 & -18.927 & -20.927 & -21.698 & -22.111 & -22.460 \\
...  & 19.929 & 19.928 & -17.603 & -18.599 & -18.827 & -18.861 & -19.062 & -18.423 & -19.419 & -19.647 & -19.681 & -19.882 \\
...  & 16.421 & 16.121 & -17.097 & -18.416 & -19.032 & -19.375 & -19.697 & -18.317 & -19.636 & -20.252 & -20.595 & -20.917 \\
\hline
\end{tabular}
\end{table}
\end{landscape}

\clearpage
\onecolumn
\tiny

\longtab{3}{ 
\begin{longtable}{|l|l|l|l|l|} 
\caption{Description of each item of the SFH catalog, of which we give an example in Table \ref{tab:sfh}. 
The five columns contain, respectively: the column ID in the catalog, the item's name as it appears in the 
database, its format, physical units and description.}
\label{tab:header}\\
\hline
\multicolumn{1}{|l|}{\sc{col}} & \multicolumn{1}{|l|}{\sc{identifier}} & \multicolumn{1}{|l|}{\sc{type}} & \multicolumn{1}{|l|}{\sc{units}} & \multicolumn{1}{|l|}{\sc{description}} \\ 
\hline
   1 & ID		 &  CHAR(25)	&  NULL 	 & WINGS identifier  \\
   2 & NAME\_SPE	 &  CHAR(18)	&  NULL 	 & File name and aperture number of the spectrum  \\
   3 & LUM\_DIST	 &  FLOAT(7,2)  &  [Mpc]	 & Luminosity distance (H0=70)  \\
   4 & CHI2\_Z05	 &  FLOAT(8,3)  &  NULL 	 & chi\^\ 2 of the best fit model with Z=0.05  \\
   5 & CHI2\_Z02	 &  FLOAT(8,3)  &  NULL 	 & chi\^\ 2 of the best fit model with Z=0.02  \\      
   6 & CHI2\_Z004	 &  FLOAT(8,3)  &  NULL 	 & chi\^\ 2 of the best fit model with Z=0.004  \\  
   7 & METAL		 &  FLOAT(5.3)  &  NULL 	 & metallicity value of the best fit model  \\
   8 & AV\_YOUNG	 &  FLOAT(7,3)  &  [mag]	 & V-band extinction, from model, of young (age bin n.1) stars  \\        
   9 & AV\_YOUNG\_ERR	 &  FLOAT(7,3)  &  [mag]	 & Uncertainty on V-band extinction, from model, of young (age bin n.1) stars  \\	
  10 & AV\_TOT  	 &  FLOAT(7,3)  &  [mag]	 & Total V-band extinction, from the model  \\  				      
  11 & AV\_TOT\_ERR	 &  FLOAT(7,3)  &  [mag]	 & Uncertainty on total V-band extinction, from the model  \\			      
  12 & SFR1		 &  FLOAT(9,4)  &[Msol.yr\^\ -1]& Star Formation Rate in the 0-2e7 yrs range  \\
  13 & SFR1\_ERR	 &  FLOAT(9,4)  &[Msol.yr\^\ -1]& Uncertainty on the Star Formation Rate in the 0-2e7 yrs range   \\  
  14 & SFR2		 &  FLOAT(9,4)  &[Msol.yr\^\ -1]& Star Formation Rate in the 2e7-6e8 yrs range   \\
  15 & SFR2\_ERR	 &  FLOAT(9,4)  &[Msol.yr\^\ -1]& Uncertainty on the Star Formation Rate in the 2e7-6e8 yrs range   \\
  16 & SFR3		 &  FLOAT(9,4)  &[Msol.yr\^\ -1]& Star Formation Rate in the 6e8-5.6e9 yrs range   \\
  17 & SFR3\_ERR	 &  FLOAT(9,4)  &[Msol.yr\^\ -1]& Uncertainty on the Star Formation Rate in the 6e8-5.6e9 yrs range \\
  18 & SFR4		 &  FLOAT(9,4)  &[Msol.yr\^\ -1]& Star Formation Rate in the 5.6e9-17.8e9 yrs range \\
  19 & SFR4\_ERR	 &  FLOAT(9,4)  &[Msol.yr\^\ -1]& Uncertainty on the Star Formation Rate in the 5.6e9-17.8e9 yrs range   \\
  20 & MASS1\_1 	 &  FLOAT(7,4)  &  NULL 	 & Percentage of stellar mass (definition n.1) in the 0-2e7 yrs range	\\	 
  21 & MASS1\_2 	 &  FLOAT(7,4)  &  NULL 	 & Percentage of stellar mass (definition n.2) in the 0-2e7 yrs range	\\	 
  22 & MASS1\_3 	 &  FLOAT(7,4)  &  NULL 	 & Percentage of stellar mass (definition n.3) in the 0-2e7 yrs range	\\	 
  23 & MASS1\_ERR	 &  FLOAT(7,4)  &  NULL 	 & Uncertainty on the percentage of stellar mass in the 0-2e7 yrs range   \\ 
  24 & MASS2\_1 	 &  FLOAT(7,4)  &  NULL 	 & Percentage of stellar mass (definition n.1) in the 2e7-6e8 yrs range   \\	 
  25 & MASS2\_2 	 &  FLOAT(7,4)  &  NULL 	 & Percentage of stellar mass (definition n.2) in the 2e7-6e8 yrs range   \\	 
  26 & MASS2\_3 	 &  FLOAT(7,4)  &  NULL 	 & Percentage of stellar mass (definition n.3) in the 2e7-6e8 yrs range   \\	 
  27 & MASS2\_ERR	 &  FLOAT(7,4)  &  NULL 	 & Uncertainty on the percentage of stellar mass in the 2e7-6e8 yrs range   \\
  28 & MASS3\_1 	 &  FLOAT(7,4)  &  NULL 	 & Percentage of stellar mass (definition n.1) in the 6e8-5.6e9 yrs range   \\      
  29 & MASS3\_2 	 &  FLOAT(7,4)  &  NULL 	 & Percentage of stellar mass (definition n.2) in the 6e8-5.6e9 yrs range   \\  	
  30 & MASS3\_3 	 &  FLOAT(7,4)  &  NULL 	 & Percentage of stellar mass (definition n.3) in the 6e8-5.6e9 yrs range   \\  	
  31 & MASS3\_ERR	 &  FLOAT(7,4)  &  NULL 	 & Uncertainty on the percentage of stellar mass in the 6e8-5.6e9 yrs range   \\
  32 & MASS4\_1 	 &  FLOAT(7,4)  &  NULL 	 & Percentage of stellar mass (definition n.1) in the 5.6e9-17.8e9 yrs range   \\      
  33 & MASS4\_2 	 &  FLOAT(7,4)  &  NULL 	 & Percentage of stellar mass (definition n.2) in the 5.6e9-17.8e9 yrs range   \\      
  34 & MASS4\_3 	 &  FLOAT(7,4)  &  NULL 	 & Percentage of stellar mass (definition n.3) in the 5.6e9-17.8e9 yrs range   \\      
  35 & MASS4\_ERR	 &  FLOAT(7,4)  &  NULL 	 & Uncertainty on the percentage of stellar mass in the 5.6e9-17.8e9 yrs range   \\
  36 & MASS\_1\_FIBER	 &  FLOAT(7,4)  &  [Msol]	 & Log10 of stellar mass (definition n.1) within the fiber aperture   \\
  37 & MASS\_2\_FIBER	 &  FLOAT(7,4)  &  [Msol]	 & Log10 of luminous stellar mass (definition n.2) within the fiber aperture  \\
  38 & MASS\_3\_FIBER	 &  FLOAT(7,4)  &  [Msol]	 & Log10 of luminous stellar mass (definition n.3) within the fiber aperture 	\\
  39 & MASS\_FIBER\_ERR  &  FLOAT(7,4)  &  [Msol]	 & Log10 of the Uncertainty on the stellar mass within the fiber aperture   \\
  40 & MASS\_1\_TOT	 &  FLOAT(7,4)  &  [Msol]	 & Log10 of total stellar mass (definition n.1)  \\
  41 & MASS\_2\_TOT	 &  FLOAT(7,4)  &  [Msol]	 & Log10 of total luminous stellar mass (definition n.2)  \\
  42 & MASS\_3\_TOT	 &  FLOAT(7,4)  &  [Msol]	 & Log10 of total luminous stellar mass (definition n.3)  \\
  43 & MASS\_TOT\_ERR	 &  FLOAT(7,4)  &  [Msol]	 & Log10 of the Uncertainty on the total stellar mass  \\
  44 & AMASS\_BJ	 &  FLOAT(7,4)  &  [Msol]	 & Log10 of the stellar mass (fiber) computed according to Bell \& DeJong (2001) \\
  45 & TMASS\_BJ	 &  FLOAT(7,4)  &  [Msol]	 & Log10 of the total stellar mass computed according to Bell \& DeJong (2001)  \\
  46 & CCOL		 &  FLOAT(7,4)  &  NULL          & Colour-aperture correction for colour gradients to the total mass  \\
  47 & LUM\_AGE\_V	 &  FLOAT(9,5)  &  [yr]          & Log10 of the V-band luminosity-weighted age  \\
  48 & LUM\_AGE 	 &  FLOAT(9,5)  &  [yr]          & Log10 of the luminosity-weighted age  \\
  49 & LUM\_AGE\_ERR	 &  FLOAT(9,5)  &  [yr]          & Uncertainty on the logarithm of the luminosity-weighted age  \\
  50 & MASS\_AGE	 &  FLOAT(9,5)  &  [yr]          & Log10 of the mass-weighted age  \\
  51 & MASS\_AGE\_ERR	 &  FLOAT(9,5)  &  [yr]          & Uncertainty on the logarithm of the mass-weighted age  \\
  52 & V\_PHOT\_FIB	 &  FLOAT(8,3)  &  [mag]         & Observed apparent V-band magnitude within the fiber   \\    
  53 & B\_PHOT\_FIB	 &  FLOAT(8,3)  &  [mag]         & Observed apparent B-band magnitude within the fiber  \\     
  54 & V\_PHOT\_TOT	 &  FLOAT(8,3)  &  [mag]         & Observed total apparent V-band magnitude  \\  
  55 & B\_PHOT\_TOT	 &  FLOAT(8,3)  &  [mag]         & Observed total apparent B-band magnitude  \\
  56 & V\_OBS\_FIB	 &  FLOAT(8,3)  &  [mag]         & Absolute V-band magnitude within the fiber, from observed spectrum  \\
  57 & B\_OBS\_FIB	 &  FLOAT(8,3)  &  [mag]         & Absolute B-band magnitude within the fiber, from observed spectrum  \\
  58 & V\_OBS\_TOT	 &  FLOAT(8,3)  &  [mag]         & Absolute total V-band magnitude from observed spectrum  \\ 
  59 & B\_OBS\_TOT	 &  FLOAT(8,3)  &  [mag]         & Absolute total B-band magnitude from observed spectrum  \\
  60 & U\_MOD\_FIB	 &  FLOAT(8,3)  &  [mag]         & Predicted observed U-band magnitude within the fiber, from the model   \\
  61 & B\_MOD\_FIB	 &  FLOAT(8,3)  &  [mag]         & Predicted observed B-band magnitude within the fiber, from the model   \\
  62 & V\_MOD\_FIB	 &  FLOAT(8,3)  &  [mag]         & Predicted observed V-band magnitude within the fiber, from the model   \\
  63 & R\_MOD\_FIB	 &  FLOAT(8,3)  &  [mag]         & Predicted observed R-band magnitude within the fiber, from the model   \\
  64 & I\_MOD\_FIB	 &  FLOAT(8,3)  &  [mag]         & Predicted observed I-band magnitude within the fiber, from the model   \\ 
  65 & J\_MOD\_FIB	 &  FLOAT(8,3)  &  [mag]         & Predicted observed J-band magnitude within the fiber, from the model   \\ 
  66 & H\_MOD\_FIB	 &  FLOAT(8,3)  &  [mag]         & Predicted observed H-band magnitude within the fiber, from the model   \\
  67 & K\_MOD\_FIB	 &  FLOAT(8,3)  &  [mag]         & Predicted observed K-band magnitude within the fiber, from the model   \\ 
  68 & U\_MOD\_TOT	 &  FLOAT(8,3)  &  [mag]         & Predicted observed total U-band magnitude, from the model   \\ 
  69 & B\_MOD\_TOT	 &  FLOAT(8,3)  &  [mag]         & Predicted observed total B-band magnitude, from the model   \\  
  70 & V\_MOD\_TOT	 &  FLOAT(8,3)  &  [mag]         & Predicted observed total V-band magnitude, from the model   \\  
  71 & R\_MOD\_TOT	 &  FLOAT(8,3)  &  [mag]         & Predicted observed total R-band magnitude, from the model   \\  
  72 & I\_MOD\_TOT	 &  FLOAT(8,3)  &  [mag]         & Predicted observed total I-band magnitude, from the model   \\  
  73 & J\_MOD\_TOT	 &  FLOAT(8,3)  &  [mag]         & Predicted observed total J-band magnitude, from the model   \\  
  74 & H\_MOD\_TOT	 &  FLOAT(8,3)  &  [mag]         & Predicted observed total H-band magnitude, from the model   \\  
  75 & K\_MOD\_TOT	 &  FLOAT(8,3)  &  [mag]         & Predicted observed total K-band magnitude, from the model   \\  
  76 & U\_ABS\_FIB	 &  FLOAT(8,3)  &  [mag]         & Predicted absolute U-band magnitude within the fiber, from the model   \\ 
  77 & B\_ABS\_FIB	 &  FLOAT(8,3)  &  [mag]         & Predicted absolute B-band magnitude within the fiber, from the model   \\  
  78 & V\_ABS\_FIB	 &  FLOAT(8,3)  &  [mag]         & Predicted absolute V-band magnitude within the fiber, from the model   \\  
  79 & R\_ABS\_FIB	 &  FLOAT(8,3)  &  [mag]         & Predicted absolute R-band magnitude within the fiber, from the model   \\  
  80 & I\_ABS\_FIB	 &  FLOAT(8,3)  &  [mag]         & Predicted absolute I-band magnitude within the fiber, from the model   \\  
  81 & J\_ABS\_FIB	 &  FLOAT(8,3)  &  [mag]         & Predicted absolute J-band magnitude within the fiber, from the model   \\  
  82 & H\_ABS\_FIB	 &  FLOAT(8,3)  &  [mag]         & Predicted absolute H-band magnitude within the fiber, from the model   \\  
  83 & K\_ABS\_FIB	 &  FLOAT(8,3)  &  [mag]         & Predicted absolute K-band magnitude within the fiber, from the model   \\  
  84 & U\_ABS\_TOT	 &  FLOAT(8,3)  &  [mag]         & Predicted absolute total U-band magnitude, from the model   \\
  85 & B\_ABS\_TOT	 &  FLOAT(8,3)  &  [mag]         & Predicted absolute total B-band magnitude, from the model   \\ 
  86 & V\_ABS\_TOT	 &  FLOAT(8,3)  &  [mag]         & Predicted absolute total V-band magnitude, from the model   \\ 
  87 & R\_ABS\_TOT	 &  FLOAT(8,3)  &  [mag]         & Predicted absolute total R-band magnitude, from the model   \\ 
  88 & I\_ABS\_TOT	 &  FLOAT(8,3)  &  [mag]         & Predicted absolute total I-band magnitude, from the model   \\
  89 & J\_ABS\_TOT	 &  FLOAT(8,3)  &  [mag]         & Predicted absolute total J-band magnitude, from the model   \\
  90 & H\_ABS\_TOT	 &  FLOAT(8,3)  &  [mag]         & Predicted absolute total H-band magnitude, from the model   \\
  91 & K\_ABS\_TOT	 &  FLOAT(8,3)  &  [mag]         & Predicted absolute total K-band magnitude, from the model   \\
  92 & usdss\_MOD\_FIB	 &  FLOAT(8,3)  &  [mag]         & Predicted observed u sdss-band magnitude within the fiber, from the model   \\
  93 & gsdss\_MOD\_FIB	 &  FLOAT(8,3)  &  [mag]         & Predicted observed g sdss-band magnitude within the fiber, from the model   \\
  94 & rsdss\_MOD\_FIB	 &  FLOAT(8,3)  &  [mag]         & Predicted observed r sdss-band magnitude within the fiber, from the model   \\
  95 & isdss\_MOD\_FIB	 &  FLOAT(8,3)  &  [mag]         & Predicted observed i sdss-band magnitude within the fiber, from the model   \\
  96 & zsdss\_MOD\_FIB	 &  FLOAT(8,3)  &  [mag]         & Predicted observed z sdss-band magnitude within the fiber, from the model   \\
  97 & usdss\_MOD\_TOT	 &  FLOAT(8,3)  &  [mag]         & Predicted observed total u sdss-band magnitude, from the model   \\
  98 & gsdss\_MOD\_TOT	 &  FLOAT(8,3)  &  [mag]         & Predicted observed total g sdss-band magnitude, from the model   \\
  99 & rsdss\_MOD\_TOT	 &  FLOAT(8,3)  &  [mag]         & Predicted observed total r sdss-band magnitude, from the model   \\
 100 & isdss\_MOD\_TOT	 &  FLOAT(8,3)  &  [mag]         & Predicted observed total i sdss-band magnitude, from the model   \\
 101 & zsdss\_MOD\_TOT	 &  FLOAT(8,3)  &  [mag]         & Predicted observed total z sdss-band magnitude, from the model   \\
 102 & usdss\_ABS\_FIB	 &  FLOAT(8,3)  &  [mag]         & Predicted absolute u sdss-band magnitude within the fiber, from the model   \\
 103 & gsdss\_ABS\_FIB	 &  FLOAT(8,3)  &  [mag]         & Predicted absolute g sdss-band magnitude within the fiber, from the model   \\
 104 & rsdss\_ABS\_FIB	 &  FLOAT(8,3)  &  [mag]         & Predicted absolute r sdss-band magnitude within the fiber, from the model   \\
 105 & isdss\_ABS\_FIB	 &  FLOAT(8,3)  &  [mag]         & Predicted absolute i sdss-band magnitude within the fiber, from the model   \\
 106 & zsdss\_ABS\_FIB	 &  FLOAT(8,3)  &  [mag]         & Predicted absolute z sdss-band magnitude within the fiber, from the model   \\
 107 & usdss\_ABS\_TOT	 &  FLOAT(8,3)  &  [mag]         & Predicted absolute total u sdss-band magnitude, from the model   \\
 108 & gsdss\_ABS\_TOT	 &  FLOAT(8,3)  &  [mag]         & Predicted absolute total g sdss-band magnitude, from the model   \\
 109 & rsdss\_ABS\_TOT	 &  FLOAT(8,3)  &  [mag]         & Predicted absolute total r sdss-band magnitude, from the model   \\
 110 & isdss\_ABS\_TOT	 &  FLOAT(8,3)  &  [mag]         & Predicted absolute total i sdss-band magnitude, from the model   \\
 111 & zsdss\_ABS\_TOT	 &  FLOAT(8,3)  &  [mag]         & Predicted absolute total z sdss-band magnitude, from the model   \\  
\hline
\end{longtable}
}
\end{document}